\documentstyle[12pt]{article}
\begin{document}

\title{
RENORMALIZATION OF POINCARE TRANSFORMATIONS IN
HAMILTONIAN SEMICLASSICAL FIELD THEORY
}
\author{
  O.Yu. Shvedov
\\
{\small {\em Sub-Dept. of Quantum Statistics and Field Theory,}}\\
{\small{\em Department of Physics, Moscow State University }}\\
{\small{\em Vorobievy gory, Moscow 119899, Russia}} }

\maketitle

\footnotetext{e-mail:
shvedov@qs.phys.msu.su}

hep-th/0109142

\begin{abstract}

Semiclassical Hamiltonian   field  theory  is  investigated  from  the
axiomatic point  of  view.  A  notion  of  a  semiclassical  state  is
introduced.  An "elementary" semiclassical state is specified by a set
of classical field configuration and quantum state  in  this  external
field. "Composed" semiclassical states viewed as formal superpositions
of "elementary" states are nontrivial only  if  the  Maslov  isotropic
condition is satisfied;  the inner product of "composed" semiclassical
states is degenerate. The mathematical proof of Poincare invariance of
semiclassical field theory is obtained for "elementary" and "composed"
semiclassical states. The notion of semiclassical field is introduced;
its Poincare invariance is also mathematically proved.

\end{abstract}

\def\qp{
\mathrel{\mathop{\bf x}\limits^2},
\mathrel{\mathop{-i\frac{\partial}{\partial {\bf x}}}\limits^1} 
}

PACS numbers: 11.30.Cp, 11.10.Cd, 11.15.Kc, 11.10.Gh.

\newcounter{eqn}[section]
\renewcommand{\theeqn}{\thesection.\arabic{eqn}}
\def\lab{\refstepcounter{eqn}\eqno(\thesection.\arabic{eqn})}
\def\l#1{\lab\label{#1}}
\def\r#1{(\ref{#1})}
\def\c#1{\cite{#1}}
\def\i#1{\bibitem{#1}}
\def\be{$$}
\def\ee{$$}
\def\bea#1 \eea{$$ \begin{array}{c} #1 \end{array} $$}
\def\bea#1 \l#2 \eea{$$ \begin{array}{c} #1 \end{array} \l{#2} $$}
\def\beb#1 \eeb{$$ \begin{array}{c} #1 \end{array} $$}
\def\crr{\\}
\sloppy

\newpage

\section{Introduction}

Different approaches to semiclassical  field  theory  have been
developed.
Most of them were based on the functional integral technique: physical
quantities were  expressed  via  functional   integrals   which   were
evaluated with the help of saddle-point or stationary-phase technique.
Since energy spectrum and $S$-matrix elements can be  found  from  the
functional integral \c{DHN,R}, this approach appeared to be useful for
the soliton quantization theory \c{DHN,R,J2,J,FK}.

Another important partial case of the semiclassical  field  theory  is
the theory  of  quantization in a strong external background classical
field \c{GMM} or in  curved  space-time  \c{BD}:  one  decomposes  the
field as  a  sum  of  a  classical  c-number  component  and a quantum
component. Then the theory is quantized.

The one-loop  approximation  \c{B1,B2,B11,B12},   the   time-dependent
Hartree-Fock   approximation   \c{B1,B2,HF1,HF2}   and  the  Gaussian
approximation developed in  \c{G1,G2,G3,G4}  may  be  also  viewed  as
examples of applications of semiclassical conceptions.

On the  other  hand,  the axiomatic field theory \c{A1,A2,A3} tells us
that main objects of QFT are  states  and  observables.  The  Poincare
group is  represented  in the Hilbert state space,  so that evolution,
boosts and  other  Poincare  transformations  are  viewed  as  unitary
operators.

The purpose of this paper is to introduce the semiclassical analogs of
such QFT notions as states,  fields and Poincare transformations.  The
analogs of  Wightman  Poincare invariance and field
axioms for the semiclassical field theory are to
be formulated and checked.

Unfortunately, "exact"  QFT  is  mathematically  constructed   for   a
restricted class     of     models    only    (see,    for    example,
\c{H,GJ,Ar1,Ar2}). Therefore,  formal  approximate  methods  such   as
perturbation theory  seem  to  be  ways  to  quantize the field theory
rather than to construct approximations for the exact solutions of QFT
equations. The   conception   of   field   quantization   within   the
perturbation framework is popular \c{BS,SF}.  One can expect that  the
semiclassical approximation plays an analogous role.

To construct  the  semiclassical  formalism  based  on the notion of a
state, one should use the equation-of-motion formulation of QFT rather
than the   usual   $S$-matrix   formulation.  It  is  well-known  that
additional difficulties such as Stueckelberg divergences  \c{Stu}  and
problems associated with the
Haag theorem  \c{Haag,A2,A3} arise in the equation-of-motion approach.
There are some ways to overcome them.  The vacuum divergences  can  be
eliminated in  the  perturbation  theory  with the help of the Faddeev
transformation \c{F}. Stueckelberg divergences can  be  treated
analogously \c{MS-F}   (exactly   solvable  models  with  Stueckelberg
divergences have   been   suggested   recently   \c{Sh1,Sh2}).   These
investigations are  important  for the semiclassical Hamiltonian field
theory \c{MS-FT}.

The semiclassical approaches are formally applicable  to  the  quantum
field theory  models if the Lagrangian depends on the fields $\varphi$
and the small parameter $\lambda$ as follows (see, for example, \c{J}):
\be
{\cal L} = \frac{1}{2} \partial_{\mu} \varphi \partial_{\mu} \varphi -
\frac{m^2}{2} \varphi^2  -  \frac{1}{\lambda}  V (\sqrt{\lambda}
\varphi),
\l{0}
\ee
where $V$ is an interaction potential.
To illustrate  the  {\it  formal}  semiclassical  ansatz for the state
vector, use  the  functional  Schrodinger  representation  (see,   for
example, \c{HF1,HF2,G3,G4}). States at fixed
moment  of  time are represented as functionals $\psi[\varphi(\cdot)]$
depending on fields $\varphi({\bf x})$,  ${\bf x} \in {\bf R}^d$,  the
field   operator   $\hat{\varphi}({\bf   x})$   is   the  operator  of
multiplication by $\varphi({\bf x})$, while the canonically conjugated
momentum  $\hat{\pi}({\bf  x})$  is  represented  as a differentiation
operator $-i\delta/\delta    \varphi({\bf    x})$.    The   functional
Schrodinger equation reads
\be
i\frac{d\psi^t}{dt} = {\cal H} \psi^t,
\l{1}
\ee
where
$$
{\cal H} = \int d{\bf x} \left[
-\frac{1}{2} \frac{\delta^2}{\delta     \varphi({\bf     x})    \delta
\varphi({\bf x})}  +  \frac{1}{2}  (\nabla   \varphi)^2({\bf   x})   +
\frac{m^2}{2} \varphi^2({\bf x}) +
\frac{1}{\lambda} V(\sqrt{\lambda} \varphi({\bf x}))
\right]
$$
The simplest semiclassical state corresponds to the Maslov  theory  of
complex germ  in  a  point  \c{M1,M2,MS3}.  It  depends  on  the small
parameter $\lambda$ as
\be
\psi^t[\varphi(\cdot)] = e^{\frac{i}{\lambda}S^t} e^{\frac{i}{\lambda}
\int d{\bf  x}  \Pi^t({\bf  x})  [\varphi({\bf   x})\sqrt{\lambda}   -
\Phi^t({\bf x})]}                 f^t\left(\varphi(\cdot)                 -
\frac{\Phi^t(\cdot)}{\sqrt{\lambda}}\right)
\equiv (K_{S^t,\Pi^t,\Phi^t} f^t)[\varphi(\cdot)],
\l{2}
\ee
where $S^t$,  $\Pi^t({\bf x})$, $\Phi^t({\bf x})$, $t\in {\bf R}$, ${\bf
x}\in {\bf R}^d$ are smooth real functions which rapidly damp with all
their derivatives as ${\bf x}\to\infty$,  $f^t[\phi(\cdot)]$ is a
$t$-dependent functional.

As $\lambda\to  0$,  the  substitution  \r{2} satisfies eq.\r{1}
in the leading order in  $\lambda$  if  the  following  relations  are
obeyed. First, for the "action" $S^t$ one finds,
\be
\frac{dS^t}{dt} = \int d{\bf x} [\Pi^t({\bf x}) \dot{\Phi}^t({\bf  x})
- \frac{1}{2}(\Pi^t({\bf  x}))^2  -  \frac{1}{2}  (\nabla  \Phi^t({\bf
x}))^2 - \frac{m^2}{2}  (\Phi^t({\bf  x}))^2  -  V(\Phi^t({\bf
x}))],
\l{3a}
\ee
Second, $\Pi^t$, $\Phi^t$ obeys the classical Hamiltonian system
\be
\dot{\Phi}^t = \Pi^t,
- \dot{\Pi}^t = (-\Delta + m^2) \Phi^t + V'(\Phi^t),
\l{3}
\ee
Finally, the functional
$f^t$ satisfies the functional Schrodinger equation with the quadratic
Hamiltonian
\be
i\dot{f}^t[\phi(\cdot)] = \int d{\bf x} \left[
- \frac{1}{2}    \frac{\delta^2}{\delta    \phi({\bf   x})   \delta
\phi({\bf x})}   +   \frac{1}{2}   (\nabla\phi({\bf   x}))^2   +
\frac{m^2}{2}\phi^2({\bf x})   +  \frac{1}{2}  V''(\Phi^t({\bf  x}))
\phi^2({\bf x})
\right] f^t[\phi(\cdot)].
\l{4}
\ee
There are   more   complicated   semiclassical   states   that  also
approximately satisfy the functional Schrodinger equation \r{1}. These
ansatzes correspond  to the Maslov theory of Lagrangian manifolds with
complex germs \c{M1,M2,MS3}. They are discussed in section 5.

However, the QFT divergences lead to the following difficulties.

It is not evident how one should  specify  the  class  of  possible
functionals $f$  and  introduce  the inner product on such a space via
functional integral. This class was constructed in \c{MS-FT}.
In particular, it was found when the Gaussian functional
\be
f[\phi(\cdot)] =  const  \exp(\frac{i}{2}  \int   d{\bf   x}d{\bf   y}
\phi({\bf x}) \phi({\bf y}) {\cal R}({\bf x},{\bf y}))
\l{5}
\ee
belongs to  this  class.  The condition on the quadratic form $\cal R$
which was obtained in \c{MS-FT} depends on $\Phi$, $\Pi$ and
differs from  the  analogous condition in the free theory.  This is in
agreement with  the  statement  of  \c{GM,Shir}   that   nonequivalent
representations of  the  canonical  commutation relations at different
moments of time should be considered if QFT  in  the  strong  external
field is investigated in the leading order in $\lambda$. However, this
does not lead to non-unitarity of the exact theory: the simple example
has been presented in \c{Sh2}.

Another problem  is  to formulate the semiclassical theory in terms of
the axiomatic field theory. Section 2 deals with formulation of axioms
of relativistic invariance and field for
the  semiclassical theory.  Section 3 is devoted to construction of
Poincare transformations.  In section 4 the  notion  of  semiclassical
field is  investigated.  More  complicated  semiclassical  states  are
constructed in section 5. Section 6 contains concluding remarks.

\section{Axioms of semiclassical field theory}

In the Wightman axiomatic approach the main object of QFT is a  notion
of a state space \c{A1,A2,A3}.  Formula \r{2} shows  us  that  in  the
semiclassical  field  theory a state at fixed moment of time should be
viewed as a set $(S,  \Pi(\cdot),  \Phi(\cdot),  f[\phi(\cdot)])$ of a
real number $S$, real functions $\Pi({\bf x})$, $\Phi({\bf x})$, ${\bf
x}\in {\bf R}^d$ and a functional $f[\phi(\cdot)]$  from  some  class.
This class depends on $\Pi$ and $\Phi$. Superposition of semiclassical
states $(S_1,\Pi_1,\Phi_1,f_1)$ and $(S_2,\Pi_2,\Phi_2,f_2)$ is of the
semiclassical  type  \r{2} if and only if $S_1=S_2$,  $\Phi_1=\Phi_2$,
$\Pi_1=\Pi_2$.

Thus, one  introduces  \c{Shv1,Shv2}  the  structure of a vector bundle
(called as a  "semiclassical  bundle"  in  \c{Shv2})  on  the  set  of
semiclassical states  of  the type \r{2}.  The base of the bundle being
a space of sets $(S,\Pi,\Phi)$   ("extended phase space"
\c{Shv1}) will be denoted as $\cal X$.
The fibers are classes of functionals which depend on
$\Phi$ and $\Pi$.  Making use of the result concerning  the  class  of
functionals \c{MS-FT}, one makes the bundle trivial as follows.
Consider the $\Phi$, $\Pi$- dependent mapping $V$
which defines a correspondence between functionals
$f$ and elements of the Fock space $\cal F$:
$$
V: \Psi \mapsto f,\qquad \Psi \in {\cal F}, \qquad f=f[\phi(\cdot)].
$$
as follows
(see, for example, \c{FJ}).
Let  $\tilde{\cal  R}({\bf x},{\bf y})$ be an
$\Phi$, $\Pi$  -  dependent symmetric function such that its imaginary
part is a kernel of a positively definite operator and  the  condition
of ref. \c{MS-FT} (see eq.\r{p1} of subsection 3.6) is satisfied. .
By $\hat{\cal R}$ we denote the operator with kernel
$\tilde{\cal  R}$,  while
$\hat{\Gamma}$ has a kernel $i^{-1}(\tilde{\cal R} - \tilde{\cal R}^*)$.
The vacuum vector of  the  Fock  space  corresponds  to  the  Gaussian
functional \r{5}.  The  operator  $V$  is  uniquely  defined  from the
relations
\bea
V^{-1} \phi({\bf x}) V = i(\hat{\Gamma}^{-1/2}(A^+-A^-))({\bf x}),\crr
V^{-1} \frac{1}{i} \frac{\delta}{\delta\phi({\bf x})} V = i (\hat{\cal
R} \hat{\Gamma}^{-1/2} A^+ - \hat{\cal R}^*
\hat{\Gamma}^{-1/2} A^-) ({\bf x}).
\l{6*}
\eea
Here $A^{\pm}({\bf x})$ are creation and annihilation operators in the
Fock space.

{\bf Definition 2.1.}  {\it A semiclassical state is a point
on the trivial bundle ${\cal X} \times {\cal F} \to {\cal
X}$.}

An important postulate of QFT is Poincare invariance.  This means that
a representation  of  the  Poincare group in the state space should be
specified. For each Poincare transformation of the form
\be
x'{}^{\mu} =  \Lambda^{\mu}_{\nu} x^{\nu} + a^{\mu},  \qquad \mu,\nu =
\overline{0,d}
\l{6**}
\ee
which is   denoted  as  $(a,\Lambda)$,  the  unitary  operator  ${\cal
U}_{a,\Lambda}$ should be specified. The group property
$$
{\cal U}_{(a_1,\Lambda_1)} {\cal U}_{(a_2,\Lambda_2)}
= {\cal U}_{(a_1,\Lambda_1)(a_2,\Lambda_2)}
$$
with
$$
(a_1,\Lambda_1)(a_2,\Lambda_2) =          (a_1+          \Lambda_1a_2,
\Lambda_1\Lambda_2).
$$
should be satisfied.

Formulate an analog of the Poincare invariance
axiom for   the   semiclassical  theory.  Suppose  that  the  Poincare
transformation ${\cal U}_{a,\Lambda}$ takes  any  semiclassical  state
$(X,f)$ to   a  semiclassical  state  $(\tilde{X},\tilde{f})$  in  the
leading order    in    $\lambda^{1/2}$.    Denote    $\tilde{X}     =
u_{a,\Lambda}X$, $\tilde{f} = U(u_{a,\Lambda}X \gets X) f$.

{\bf Axiom 1 (Poincare invariance)}
\\
{\it
( ) the mappings $u_{a,\Lambda}: {\cal X} \to {\cal X}$ are specified,
the group properties for them
$u_{a_1,\Lambda_1}
u_{a_2,\Lambda_2} = u_{(a_1,\Lambda_1)(a_2,\Lambda_2)}$ are satisfied;\\
(¡) for all  $X\in   {\cal   X}$  the unitary operators
$U_{a,\Lambda} (u_{a,\Lambda}X  \gets  X):  {\cal  F}  \to  {\cal F}$,
obeying the group property
\bea
U_{a_1,\Lambda_1} (u_{(a_1,\Lambda_1)(a_2,\Lambda_2)}X  \gets
u_{(a_2,\Lambda_2)}X)
U_{a_2,\Lambda_2} (u_{(a_2,\Lambda_2)}X  \gets  X)
=\crr
U_{(a_1,\Lambda_1)(a_2,\Lambda_2)}
(u_{(a_1,\Lambda_1)(a_2,\Lambda_2)}X  \gets X)
\l{axiom1}
\eea
are specified.
}

An important  feature  of QFT is the notion of a field:  it is assumed
that an  operator  distribution  $\hat{\varphi}({\bf  x},t)$
is  specified.
Investigate  it  in  the  semiclassical theory.  Applying the operator
$\varphi({\bf x})$ to the semiclassical  state  \r{2},  we  obtain  an
analogous state:
$$
e^{\frac{i}{\lambda}S^t} e^{\frac{i}{\lambda}
\int d{\bf  x}  \Pi^t({\bf  x})  [\varphi({\bf   x})\sqrt{\lambda}   -
\Phi^t({\bf x})]}
\tilde{f}^t(\varphi(\cdot)                 -
\frac{\Phi^t(\cdot)}{\sqrt{\lambda}}),
$$
where
$$
\tilde{f}^t[\phi(\cdot)] = (\lambda^{-1/2} \Phi^t({\bf x}) + \phi({\bf
x}) ) f^t[\phi(\cdot)]
$$
As $\lambda\to 0$, one has
$$
\hat{\varphi}({\bf x},t) =
\lambda^{-1/2} \Phi^t({\bf x}) + \hat{\phi}({\bf x},t:X),
$$
where $\hat{\phi}({\bf  x},t:X)$  is a $\Pi,\Phi$-dependent operator
in ${\cal F}$,  $\Phi^t({\bf x})\equiv \Phi(x:X)$ is a solution to the
Cauchy problem for eq.\r{3}.
The  field  axiom  can be reformulated as
follows.

{\bf Axiom 2}.  {\it For each $X\in  {\cal  X}$  the
operator distribution  $\phi({\bf  x},t;X):  {\cal F} \to {\cal F}$ is
specified.}

An important  feature  of the relativistic quantum field theory is the
property of Poincare invariance of fields.  The operator  distribution
$\hat{\varphi}({\bf x},t)$ should obey the following property
$$
{\cal U}_{a,\Lambda}  \hat{\varphi}(x) = \hat{\varphi} (\Lambda x + a)
{\cal U}_{a,\Lambda}.
$$
Apply this identity to  a  semiclassical  state  $(X,f)$.  In  leading
orders in $\lambda^{1/2}$, one obtains:
\beb
\lambda^{-1/2} \Phi(x:X)
(u_{a,\Lambda}X,  U_{a,\Lambda}(u_{a,\Lambda} X \gets X) f)
+
(u_{a,\Lambda}X,  U_{a,\Lambda}(u_{a,\Lambda} X \gets X) \hat{\phi}(x:X)
f)
= \crr
\lambda^{-1/2} \Phi(\Lambda x+a: u_{a,\Lambda}X)
(u_{a,\Lambda}X,  U_{a,\Lambda}(u_{a,\Lambda} X \gets X) f)
\crr
+
(u_{a,\Lambda}X,  \hat{\phi}(\Lambda x + a: u_{a,\Lambda}X)
U_{a,\Lambda}(u_{a,\Lambda} X \gets X) f).
\eeb
Therefore, we formulate the following axiom.

{\bf Axiom  3.  (Poincare  invariance  of fields).} {\it The following
properties are satisfied:
\be
\Phi(x:X) = \Phi(\Lambda x+a:u_{a,\Lambda}X);
\l{aa1}
\ee
\be
\hat{\phi}(\Lambda x + a: u_{a,\Lambda}X)
U_{a,\Lambda}(u_{a,\Lambda} X \gets X)
= U_{a,\Lambda}(u_{a,\Lambda} X \gets X)
\hat{\phi}(x: X).
\l{aa2}
\ee
}

\section{Semiclassical Poincare transformations}

\subsection{Construction of poincare transformations in the functional
representation}

{\bf 1}.
Let us construct the mappings $u_{a,\Lambda}$  and  unitary  opertaors
$U_{a,\Lambda} (u_{a,\lambda}X   \gets   X)$.   Since   any   Poincare
transformation is a composition of time and space translations,  boost
and spatial rotations,
$$
(a,\Lambda) = (a^0,0) ({\bf a},0) (0, \exp(\alpha^k l^{0k}))
(0,\exp(\frac{1}{2} \theta_{sm} l^{sm}))
$$
with $\theta_{sm}=-\theta_{ms}$,
$$
(l^{\lambda\mu})^{\alpha}_{\beta}   =   -  g^{\lambda
\alpha} \delta^{\mu}_{\beta} + g^{\mu\alpha} \delta^{\lambda}_{\beta},
$$
it is  sufficient  to  specify  operators  $U_{a,\Lambda}$  for these
special cases and then apply a group property.

In the "exact" theory, the operator
${\cal U}_{a,\Lambda}$
has the form
\be
{\cal U}_{a,\Lambda} =
\exp[i{\cal P}^{0}a^0]
\exp[-i{\cal P}^{j}a^j]
\exp[{i}\alpha^k{\cal M}^{0k}]
\exp[\frac{i}{2}{\cal M}^{lm} \theta_{lm}].
\l{b1}
\ee
The momentum and angular momentum operators entering to formula \r{b1}
have the well-known form (see,  for example, \c{BS})
\be
{\cal P}^{\mu} = \int d{\bf x} T^{\mu 0}({\bf x}),
\qquad
{\cal M}^{\mu\lambda} = \int d{\bf x} [x^{\mu} T^{\lambda 0}({\bf x})
- x^{\lambda} T^{\mu 0}({\bf x})],
\l{et1}
\ee
where formally
$$
T^{00} = \frac{1}{2} \hat{\pi}^2 + \frac{1}{2} \partial_i \hat{\varphi}
\partial_i \hat{\varphi}    +    \frac{m^2}{2}    \hat{\varphi}^2    +
\frac{1}{\lambda} V(\sqrt{\lambda} \hat{\varphi}),
\qquad
T^{k0} = - \partial_k\hat{\varphi} \hat{\pi}.
$$
We are  going  to apply the operator \r{b1} to the semiclassical state
\r{2}. Note  that  the   operators   ${\cal   P}^{\mu}$   and   ${\cal
M}^{\mu\nu}$ \r{et1}  depend  on  field  $\hat{\varphi}$  and  momentum
$\hat{\pi}$ semiclassically,
$$
{\cal P}^{\mu}    =    \frac{1}{\lambda}    P^{\mu}    (\sqrt{\lambda}
\hat{\pi}(\cdot), \sqrt{\lambda}\hat{\varphi}(\cdot)),
\qquad
{\cal M}^{\mu\nu} =    \frac{1}{\lambda}    M^{\mu\nu}    (\sqrt{\lambda}
\hat{\pi}(\cdot), \sqrt{\lambda} \hat{\varphi}(\cdot)),
$$
It is convenient to consider the more general problem  (cf.\c{M2}).Let
us find as $\lambda\to 0$ the state
\be
\exp(-i{\cal A}) K_{S^0,\Pi^0,\Phi^0} f^0,
\l{e2}
\ee
where $K_{S,\Pi,\Phi}$ has the form \r{2},
$$
{\cal A}   =    \frac{1}{\lambda}    A(\sqrt{\lambda}\hat{\pi}(\cdot),
\sqrt{\lambda} \hat{\varphi}(\cdot)).
$$
Note that  the  state functional \r{e2} may be viewed as a solution to
the Cauchy problem of the form
\bea
i \frac{\partial\Psi^{\tau}}{\partial\tau}     =     \frac{1}{\lambda}
A(\frac{\sqrt{\lambda}}{i} \frac{\delta}{\delta        \varphi(\cdot)},
\sqrt{\lambda} \varphi(\cdot)) \Psi^{\tau}, \crr
\Psi^0[\varphi(\cdot)] = (K_{S^0,\Pi^0,\Phi^0} f^0)[\varphi(\cdot)]
\l{et3}
\eea
at $\tau=1$.  Let us look for the asymptotic solution to eq.\r{et3}  in
the following form:
\be
\Psi^{\tau}[\varphi(\cdot)]                                          =
(K_{S^{\tau},\Pi^{\tau},\Phi^{\tau}}f^{\tau})[\varphi(\cdot)].
\l{e4}
\ee
Substitution of  functional \r{e4} to eq.\r{et3} gives us the following
relation:
\bea
[-\frac{1}{\lambda} (\dot{S}^{\tau}  -  \int  d{\bf x} \Pi^{\tau}({\bf
x}) \dot{\Phi}^{\tau}({\bf x})) - \frac{1}{\sqrt{\lambda}} \int  d{\bf
x} (\dot{\Pi}^{\tau}({\bf  x})  \phi({\bf x}) + \dot{\Phi}^{\tau}({\bf
x}) i    \frac{\delta}{\delta     \phi({\bf     x})}     )     +     i
\frac{\partial}{\partial\tau} ] f^{\tau}[\phi(\cdot)] =\crr
\frac{1}{\lambda} A(\Pi^{\tau}(\cdot)        -         i\sqrt{\lambda}
\frac{\delta}{\delta \phi(\cdot)}, \Phi^{\tau}(\cdot) + \sqrt{\lambda}
\phi(\cdot)) f^{\tau}[\phi(\cdot)].
\l{e5}
\eea
Considering the    terms    of    the    orders     $O(\lambda^{-1})$,
$O(\lambda^{-1/2})$ and $O(1)$ in eq.\r{e5}, we obtain
\be
\dot{S}^{\tau} =     \int     d{\bf     x}    (\Pi^{\tau}({\bf     x})
\dot{\Phi}^{\tau}({\bf x}) - A(\Pi^{\tau}(\cdot),\Phi^{\tau}(\cdot)),
\l{e6}
\ee
\be
\dot{\Phi}^{\tau}({\bf x})  =   \frac{\delta   A   (\Pi^{\tau}(\cdot),
\Phi^{\tau}(\cdot))}{\delta \Pi({\bf x})},
\qquad
\dot{\Pi}^{\tau}({\bf x})  = - \frac{\delta   A   (\Pi^{\tau}(\cdot),
\Phi^{\tau}(\cdot))}{\delta \Phi({\bf x})},
\l{e7}
\ee
\bea
i \frac{\partial f^{\tau}[\phi(\cdot)]}{\partial\tau} = \left(
\int d{\bf x} d{\bf y} \left[
\frac{1}{2} \frac{1}{i} \frac{\delta}{\delta \phi({\bf x})}
\frac{\delta^2A}{\delta\Pi({\bf x}) \delta\Pi({\bf y})}
\frac{1}{i} \frac{\delta}{\delta \phi({\bf y})}
+ \right. \right. \crr
\left. \left.
\phi({\bf x})
\frac{\delta^2A}{\delta\Phi({\bf x}) \delta\Pi({\bf y})}
\frac{1}{i} \frac{\delta}{\delta \phi({\bf y})}
+
\frac{1}{2} \phi({\bf x})
\frac{\delta^2A}{\delta\Phi({\bf x}) \delta\Phi({\bf y})}
\phi({\bf y})
\right] + A_1 \right) f^{\tau}[\phi(\cdot)].
\l{e8}
\eea
Here $A_1$ is a c-number quantity which depends on the ordering of the
operators $\hat{\varphi}$  and  $\hat{\pi}$  and  is  relevant  to the
renormalization problem.

We see that for the cases ${\cal A} = - {\cal P}^0a^0$,  ${\cal  A}  =
{\cal P}^ja^j$,  ${\cal  A}  =  -\alpha^k {\cal M}^{0k}$,  ${\cal A} =
\frac{1}{2} \theta_{sm}{\cal  M}^{sm}$  the  mapping   $u_{a,\Lambda}$
takes the  initial  condition  for  the  system \r{e6},  \r{e7} to the
solution of the Cauchy  problem  for  this  system  at  $\tau=1$.  The
operators $\tilde{U}_{a,\Lambda}$ transforms the initial condition for
eq.\r{e8} to the solution at $\tau=1$.

{\bf 2.}
The classical mappings  $u_{a,\Lambda}$  for  our  partial  cases  are
presented in table 1.

\begin{table}
\caption{Poincare transformations in classical theory}
\begin{tabular}{|p{4cm}|p{7cm}|p{7cm}|}
\hline
\parbox{4cm}{
Element of Poincare group
$(a_{\tau},\Lambda_{\tau})$
}
&
\parbox{7cm}{
Classical Poincare transformation
$$
u_{a_{\tau},\Lambda_{\tau}}: (S^0,\Pi^0,\Phi^0) \mapsto
(S^{\tau},\Pi^{\tau},\Phi^{\tau})
$$
}
&
\parbox{7cm}{
Classical Lie derivative
$$
\delta F(S,\Pi,\Phi)            =           \frac{d}{d\tau}|_{\tau=0}
F(S^{\tau},\Pi^{\tau},\Phi^{\tau})
$$
}
\\
\hline
\parbox{4cm}{
$a_{\tau}=0$, 
\\ 
$\Lambda_{\tau}           =          \exp(\frac{\tau}{2}
l^{sm}\theta_{sm})$; \\
spatial rotation
}
&
\parbox{7cm}{
$$\Phi^{\tau}({\bf x})   =   \Phi^0(e^{-\frac{\tau}{2}l^{sm}\theta_{sm}}
{\bf x});$$
$$\Pi^{\tau}({\bf x})   =   \Pi^0(e^{-\frac{\tau}{2}l^{sm}\theta_{sm}}
{\bf x});$$
$$S^{\tau} = S^0$$
}
&
\parbox{7cm}{
\beb
\frac{1}{2} \theta_{lm} \delta^{lm}_M =\\
\frac{1}{2}\theta_{lm}
\int d{\bf x}
((x^l\partial_m - x^m\partial_l) \Phi({\bf x})
\frac{\delta}{\delta \Phi({\bf x})}\\
+ (x^l\partial_m - x^m\partial_l) \Pi({\bf x})
\frac{\delta}{\delta \Pi({\bf x})})
\eeb
}
\\
\hline
\parbox{4cm}{
$a_{\tau}^0 = 0$, $\Lambda_{\tau} =1$,\\
${\bf a}_{\tau} = {\bf b}\tau$;\\
spatial translation
}
&
\parbox{7cm}{
$$
\Phi^{\tau}({\bf x}) = \Phi^0({\bf x}- {\bf b}\tau);
$$
$$
\Pi^{\tau}({\bf x}) = \Pi^0({\bf x}- {\bf b}\tau);
$$
$$
S^{\tau} = S^0.
$$
}
&
\parbox{7cm}{
\beb
-b^k \delta_P^k = \\
- b^k \int  d{\bf  x}
(\partial_k\Phi({\bf  x})  \frac{\delta}{\delta
\Phi({\bf x})} +
\partial_k\Phi({\bf  x})  \frac{\delta}{\delta
\Phi({\bf x})})
\eeb
}
\\
\hline
\parbox{4cm}{
$a^0 = - \tau$, ${\bf a}=0$;\\
$\Lambda = 1$;\\
evolution
}
&
\parbox{7cm}{
Resolving operator for the Cauchy problem:
\beb
\dot{\Phi}^{\tau} = \Pi^{\tau};\\
- \dot{\Pi}^{\tau} = (-\Delta + m^2) \Phi^{\tau} + V'(\Phi^{\tau});\\
\dot{S}^{\tau} = \int d{\bf x} [\Pi^{\tau} \dot{\Phi}^{\tau}
- \frac{1}{2}(\Pi^{\tau})^2  -\\
\frac{1}{2}  (\nabla  \Phi^{\tau})^2
- \frac{m^2}{2}  (\Phi^{\tau})^2  -  V(\Phi^{\tau})].
\eeb
}
&
\parbox{7cm}{
\beb
- \delta_H  = \\
\int d{\bf x} [
\Pi({\bf x})   \frac{\delta }{\delta  \Phi  ({\bf  x})}  -  (-\Delta
\Phi({\bf x})\\ + m^2 \Phi({\bf x}) +  V'(\Phi({\bf  x})))  \frac{\delta
}{\delta \Pi({\bf x})}
] - \\
\int   d{\bf   x}
[\frac{1}{2}
\Pi^2({\bf x})  - \frac{1}{2} (\nabla \Phi({\bf x}))^2 \\
- \frac{m^2}{2}
\Phi^2({\bf x}) - V(\Phi({\bf x}))]
\frac{\partial}{\partial  S}.
\eeb
}
\\
\hline
\parbox{4cm}{
$a_{\tau} =0$; \\
$\Lambda_{\tau} = \exp[-\tau n^kl^{0k}]$;\\
boost
}
&
\parbox{7cm}{
Resolving operator for the Cauchy problem
\beb
\dot{\Phi}^{\tau} = n^kx^k \Pi^{\tau},
\\
- \dot{\Pi}^{\tau} = - \nabla x^kn^k \nabla \Phi^{\tau}
\\
+  x^kn^k  (m^2\Phi^{\tau}  +   V'(\Phi^{\tau})),
\\
\dot{S}^{\tau} =    \int     d{\bf     x}     [\Pi^{\tau}
\dot{\Phi}^{\tau}   -  x^kn^k  [\frac{1}{2}  (\Pi^{\tau}
)^2 + \\
\frac{1}{2} (\nabla \Phi^{\tau})^2  +  \frac{m^2}{2}
(\Phi^{\tau})^2 + V(\Phi^{\tau})].
\eeb
}
&
\parbox{7cm}{
\beb
- n^m \delta_B^m  = \\
n^m \int d{\bf x} [x^m \Pi({\bf
x}) \frac{\delta }{\delta \Phi({\bf x})} -  (-\partial_ix^m\partial_i
\Phi({\bf x}) \\
+  x^m  m^2\Phi({\bf  x})  +  x^m  V'(\Phi({\bf  x})) )
\frac{\delta }{\delta \Pi({\bf x})}] + \\
n^m
\int d{\bf x} x^m
[\frac{1}{2}
\Pi^2({\bf x})  - \frac{1}{2} (\nabla \Phi({\bf x}))^2 -
\\
\frac{m^2}{2}
\Phi^2({\bf x}) - V(\Phi({\bf x}))]
\frac{\partial }{\partial S}.
\eeb
}
\\
\hline
\end{tabular}
\end{table}

One can write down the following general formula.
Let $(a,\Lambda)$be an arbitrary Poincare transformation.  It  happens
that the  mapping  $u_{a,\Lambda}:  (S,\Pi,\Phi)  \mapsto  (\tilde{S},
\tilde{\Pi}, \tilde{\Phi})$ has the  following  form.  Let  $\Phi({\bf
x},t) \equiv \Phi(x)$ be a solution of the Cauchy problem
\bea
\partial_{\mu} \partial^{\mu} \Phi(x) + m^2\Phi(x) + V{}'(\Phi(x)) =
0, \crr
\Phi({\bf x},0) = \Phi({\bf x}), \qquad
\frac{\partial}{\partial t} \Phi({\bf x},t)|_{t=0} = \Pi({\bf x}).
\l{e13}
\eea
Denote
$$
\breve{\Phi}(x) = \Phi(\Lambda^{-1}(x-a)).
$$
It appears that
\bea
\tilde{\Phi}({\bf x}) = \breve{\Phi}({\bf x},0),
\qquad
\tilde{\Pi}({\bf x}) = \frac{\partial}{\partial t}
\breve{\Phi}({\bf x},t)|_{t=0},\crr
\tilde{S} = S  +  \int  dx  [\theta(x^0)  \theta(-(\Lambda  x+a)^0)  -
\theta(-x^0) \theta((\Lambda x+a)^0)]\crr
\times
[\frac{1}{2} \partial_{\mu}   \Phi(x)   \partial^{\mu}    \Phi(x)    -
\frac{{m^2}}{2} \Phi^2(x) - V(\Phi(x)).
\l{e14}
\eea
For spatial translations,  rotations and evolution,  agreement between
\r{e14} and table 1 is evident.  Consider the $x^1$-boost case, $n^k =
(1,0,...,0)$. One has
\beb
\tilde{\Phi}_{\tau} ({\bf  x})  = \Phi(x^1 \cosh\tau + x^0
\sinh \tau,  x^2,
..., x^d, x^0 \cosh\tau + x^1 \sinh\tau)|_{x^0=0},
\crr
\tilde{\Pi}_{\tau} ({\bf  x})  = \frac{\partial}{\partial x^0}
\Phi(x^1 \cosh\tau + x^0 \sinh \tau,  x^2,
..., x^d, x^0 \cosh\tau + x^1 \sinh\tau)|_{x^0=0},
\eeb
The functions $\Phi_{\tau}$, $\Pi_{\tau}$
obey the system presented in table 1.
For the integral for $\tilde{S}$,  consider  the
substitution $x^0  =  y^1  \sinh\tilde{\tau}$,  $x^1=y^1 \cosh\tilde{\tau}$,
$x^2=y^2$,..., $x^d=y^d$. One finds
$$
\tilde{S}^{\tau} = S + \int_0^{\tau} d\tilde{\tau} y^1 d{\bf y} [
\frac{1}{2} (\tilde{\Pi}_{\tau}({\bf y}))^2 -
\frac{1}{2} (\nabla \tilde{\Phi}_{\tau}({\bf y}))^2 -
\frac{m^2}{2} \tilde{\Phi}_{\tau}^2({\bf             y})             -
V(\tilde{\Phi}_{\tau}({\bf y}))
]
$$
this agrees with table 1.

One can  also  notice  that  the  group  property  for  eq.\r{e14}  is
satisfied.

Let us make more precise the definition of the space $\cal X$.

{\bf Definition 3.1.} {\it $\cal X$ is a space of sets  $(S,\Pi,\Phi)$
of a  number  $S$  and functions $\Pi,\Phi \in S({\bf R}^d)$ such that
there exists a unique solution of the Cauchy problem \r{e13} such that
the functions $\Phi(\Lambda x+a)|_{x^0=0}$ and
$\partial_{\mu}\Phi(\Lambda x+a)|_{x^0=0}$ are of  the  class  $S({\bf
R}^d)$ for all $a.\Lambda$. }

We see that the transformation $u_{a,\Lambda}:  {\cal X} \to {\cal X}$
is defined.

{\bf 3.}
The operators  $\tilde{U}_{a,\Lambda}(u_{a,\Lambda}  X  \gets  X)$ are
presented in table 2.

\begin{table}
\caption{Semiclassical Poincare    transformations    in    functional
representation}
\begin{tabular}{|p{4cm}|p{14cm}|}
\hline
\parbox{4cm}{
Element of Poincare group
$(a_{\tau},\Lambda_{\tau})$
}
&
\parbox{14cm}{
Semiclassical operator $\tilde{U}_{a_{\tau},\Lambda_{\tau}}
(u_{a_{\tau},\Lambda_{\tau}} X \gets X) : f_0 \mapsto f_t$
in the functional representation takes the initial condition  for  the
Cauchy problem to the solution of the Cauchy problem for the equation:
}
\\
\hline
\parbox{4cm}{
$a_{\tau}=0$, 
\\ 
$\Lambda_{\tau}           =          \exp(\frac{\tau}{2}
l^{sm}\theta_{sm})$; \\
spatial rotation
}
&
\parbox{14cm}{
$$
i \dot{f}^{\tau}[\phi(\cdot)]
 = - \frac{1}{2} \theta_{sm} \tilde{M}^{sm}(X_{\tau})
 f_{\tau}[\phi(\cdot)];
$$
\beb
\tilde{M}^{sm} = - \int d{\bf x}  [(x^s  \partial_m  -  x^m\partial_s)
\phi({\bf x})] \frac{1}{i} \frac{\delta}{\delta \phi({\bf x})}.
\eeb
}
\\
\hline
\parbox{4cm}{
$a_{\tau}^0 = 0$, $\Lambda_{\tau} =1$,\\
${\bf a}_{\tau} = {\bf b}\tau$;\\
spatial translation
}
&
\parbox{14cm}{
$$
i\dot{f}^{\tau}[\phi(\cdot)] = b^k \tilde{P}^k (X_{\tau})
f_{\tau}[\phi(\cdot)];
$$
\beb
\tilde{P}^k =  -  \int  d{\bf  x}  \partial_k\phi({\bf x}) \frac{1}{i}
\frac{\delta}{\delta \phi({\bf x})}.
\eeb
}
\\
\hline
\parbox{4cm}{
$a^0 = - \tau$, ${\bf a}=0$;\\
$\Lambda = 1$;\\
evolution
}
&
\parbox{14cm}{
$$
i\dot{f}^{\tau}[\phi(\cdot)] = \tilde{H}(X_{\tau}) f_{\tau}[\phi(\cdot)];
$$
\beb
\tilde{H} = \int d{\bf x} \left[
- \frac{1}{2} \frac{\delta^2}{\delta \phi({\bf  x})  \delta  \phi({\bf
x})} +   \frac{1}{2}   (\nabla   \phi)^2({\bf   x})   +  \frac{m^2}{2}
\phi^2({\bf x}) + \frac{1}{2} V''(\Phi({\bf x})) \phi^2({\bf x}).
\right]
\eeb
}
\\
\hline
\parbox{4cm}{
$a_{\tau} =0$; \\
$\Lambda_{\tau} = \exp[-\tau n^kl^{0k}]$;\\
boost
}
&
\parbox{14cm}{
$$
i\dot{f}^{\tau}[\phi(\cdot)] = n^m \tilde{B}^m(X_{\tau}) f_{\tau}
[\phi(\cdot)];
$$
\beb
\tilde{B}^m = \int d{\bf x} x^m
\left[
- \frac{1}{2}  \frac{\delta^2}{\delta  \phi({\bf  x}) \delta \phi({\bf
x})} +  \frac{1}{2}   (\nabla   \phi)^2({\bf   x})   +   \frac{m^2}{2}
\phi^2({\bf x}) + \frac{1}{2} V''(\Phi({\bf x})) \phi^2({\bf x})
\right].
\eeb
}
\\
\hline
\end{tabular}
\end{table}

However, it is not easy to check the group property \r{axiom1}.  It is
much more  convenient  to  investigate  the   infinitesimal   Poincare
transformations and check the algebraic analog of \r{axiom1}.

It happens that operators $\tilde{U}_{a,\Lambda} (u_{a,\Lambda}X \gets
X)$ induce a Poincare group representation in a specific space.  It is
a space  of  sections  $f(x;\phi(\cdot))$ of the semiclassical bundle.
The operators $\breve{\tilde{U}}_{a,\Lambda}$ act as
\be
(\breve{\tilde{U}}_{a,\Lambda} f)(X)  =  \tilde{U}_{a,\Lambda}(X\gets
u_{a,\Lambda}^{-1} X) f(u_{a,\Lambda}^{-1} X).
\l{repres}
\ee
The group property for the operators $\breve{\tilde{U}}$ is equivalent
to relation    \r{axiom1}.    Let    $(a_{tau},\Lambda_{\tau})$  be  a
one-parametric subgroup of the Poincare group with the tangent  vector
$A$ being  an  element  of  the  Poincare algebra.  Since the operator
$\tilde{U}_{a_{\tau},\Lambda_{\tau}} (u_{a_{\tau},\Lambda_{\tau}}    X
\gets X)$  takes  the  initial  condition  for  the cauchy problem for
equation
$$
i\dot{f}_{\tau} = \tilde{H}(A:u_{a_{\tau},\Lambda_{\tau}}X) f_{\tau}
$$
to the  solution  of  this  equation.  therefore,  the  generator   of
representation \r{repres} is
$$
(\breve{\tilde{H}} (A) f) (X) = i \frac{d}{d\tau}|_{\tau = 0}
(\breve{\tilde{U}}_{a_{\tau},\Lambda_{\tau}} f)(X) =
[\tilde{H}(A:X) - i\delta[A]] f(X),
$$
where
$$
\delta[A] = \frac{d}{d\tau}|_{\tau=0} f(u_{a_{\tau},\Lambda_{\tau}} X)
$$
is a Lie derivative presented in table 1. Therefore, the infinitesimal
analog of the group property \r{axiom1} is
\be
[\tilde{H}(A_1:X) - i\delta[A_1];
\tilde{H}(A_2:X) - i\delta[A_2]] =
i (\tilde{H}([A_1;A_2]:X) - i\delta[A_1;A_2]).
\l{alg1}
\ee
It follows from notations of tables 1 and 2 that relation \r{alg1} can
be rewritten for the Poincare algebra as
$$
[\breve{\tilde{P}}^{\lambda}, \breve{\tilde{P}}^{\mu}] = 0;\qquad
[\breve{\tilde{M}}^{\lambda \mu}, \breve{\tilde{P}}^{\sigma}]
=  i(g^{\mu \sigma}\breve{\tilde{P}}^{\lambda}    -    g^{\lambda \sigma}
\breve{\tilde{P}}^{\mu})
$$
\be
[\breve{M}^{\lambda\mu}, \breve{M}^{\rho\sigma}]
=    -i(g^{\lambda \rho}\breve{M}^{\mu\sigma}     -
g^{\mu\rho}\breve{M}^{\lambda \sigma}
+      g^{\mu\sigma}     \breve{M}^{\lambda \rho}     -
g^{\lambda\sigma} \breve{M}^{\mu\rho}).
\l{f5}
\ee
for operators
$$
\breve{\tilde{M}}^{ms} = \tilde{M}^{ms} + i \delta_M^{ms},
\qquad
\breve{\tilde{P}}^{m} = \tilde{P}^{m} + i \delta_P^{m},
\qquad
\breve{\tilde{P}}^{0} = \tilde{H} + i \delta_H,
\qquad
\breve{\tilde{M}}^{k0} = \tilde{B}^{k} + i \delta_B^{k}
$$
It is  checked  by  direct  calculations  that  eqs.\r{f5}  are   {\it
formally}
staisfied. However,   there   is   a   problem   of   divergences  and
renormalization which requires more careful investigations.

\subsection{Semiclassical Poincare transformations in Fock space}

For renormalization, let  us  construct  the   semiclassical   Poincare
transformations in  the  Fock  space.  They  are  related   with   the
constructed operators  $\tilde{U}_{a,\Lambda}(u_{a,\Lambda}X \gets X)$
by the relation:
\be
\tilde{U}_{a,\Lambda} (u_{a,\Lambda}X \gets X) =
V_{u_{a,\Lambda}X} U_{a,\Lambda} (u_{a,\Lambda}X \gets X) V_X^{-1}.
\l{b0}
\ee
The operator $V$ taking the Fock space vector $\Psi\in {\cal
F}$ to the functional $f[\phi(\cdot)]$ is defined from the relation
\be
V: |0>  \mapsto  c\exp[\frac{i}{2}\int d{\bf x} d{\bf y}
\tilde{\cal R}({\bf x},{\bf y}) \phi({\bf x}) \phi({\bf y})]
\l{b11}
\ee
and from formulas \r{6*} which can be rewritten as
\bea
VA^{+}({\bf x}) V^{-1} = {\cal A}^{+}({\bf x}) \equiv
(\hat{\Gamma}^{-1/2}\hat{\cal R}^*\phi
-      \hat{\Gamma}^{-1/2}     \frac{1}{i}
\frac{\delta}{\delta\phi}) ({\bf x}), \crr
VA^{-}({\bf x}) V^{-1} = {\cal A}^{-}({\bf x}) \equiv
(\hat{\Gamma}^{-1/2}\hat{\cal R}\phi
-      \hat{\Gamma}^{-1/2}     \frac{1}{i}
\frac{\delta}{\delta\phi}) ({\bf x}).
\l{b12}
\eea
$|c|$  can be formally found from the normalization condition
\be
|c|^2 \int D\phi |\exp[\frac{i}{2}\int d{\bf x} d{\bf y} \phi({\bf x})
\tilde{\cal R}({\bf x},{\bf y}) \phi({\bf y})]|^2 = 1
\l{b13}
\ee
The argument can be chosen to be arbitrary, for example,
\be
Arg c = 0.
\l{b14}
\ee

Notice that the operator  $V$ is defined form
the relations \r{b11} - \r{b14} uniquely.

Namely, any element of the Fock space can be presented
\c{Ber} via its components, vacuum state an creation operators as
$$
\Psi = \sum_{n=0}^{\infty} \frac{1}{\sqrt{n!}}  \int  d{\bf  x}_1  ...
d{\bf x}_n \Psi_n({\bf x}_1,...,{\bf x}_n) A^+({\bf x}_1) ... A^+({\bf
x}_n) |0>
$$
Specify 
$$
V\Psi = \sum_{n=0}^{\infty} \frac{1}{\sqrt{n!}}  \int  d{\bf  x}_1  ...
d{\bf x}_n  \Psi_n({\bf  x}_1,...,{\bf x}_n) {\cal A}^+({\bf x}_1) ...
{\cal A}^+({\bf
x}_n) V|0>.
$$
{The problem of divergence of the series is related with
the problem   of   correctness   of   the    functional    Schrodinger
representation. It is not investigated here.}

Since the  operators ${\cal A}^{\pm}({\bf x})$ satisfy usual canonical
commutation relations and
${\cal A}^-({\bf x})|0>=0$,
we obtain $VA^{\pm}({\bf x}) = {\cal A}^{\pm}({\bf x})V$.

The operator $V$  depend on  $\cal  R$.  It is useful to find
an explicit form of the operator
$V^{-1}\delta V$.

It happens that the following property is satisfied:
\bea
V^{-1}\delta V = - \frac{i}{2} A^+ \hat{\Gamma}^{-1/2} \delta \hat{\cal R}
\hat{\Gamma}^{-1/2} A^+ - \frac{i}{2} A^- \hat{\Gamma}^{-1/2}
\delta  \hat{\cal  R}^*
\hat{\Gamma}^{-1/2} A^-   +
\crr
  A^+   [\hat{\Gamma}^{1/2}   \delta  \hat{\Gamma}^{-1/2}  +
i\hat{\Gamma}^{-1/2} \delta \hat{\cal R}^* \hat{\Gamma}^{-1/2}] A^-
+  \frac{i}{4}  Tr
[\delta(\hat{\cal R} +  \hat{\cal R}^*) \hat{\Gamma}^{-1}]
\l{b15}
\eea

The notations of the type
$A^+\hat{\cal  B}A^-$  are used for the operators like
$\int  d{\bf  x}  d{\bf y} A^+({\bf x}) \tilde{\cal B}({\bf
x},{\bf y}) A^-({\bf y}) $,  where $\tilde{\cal B}({\bf x},{\bf  y})$  is  a
kernel of the operator $\hat{\cal B}$.

To check formula \r{b15}, consider the variation of the formula
\r{6*} if $\cal R$ is varied:
\beb
[A^{\pm}({\bf x})   ;   V^{-1}\delta   V]   =   (\hat{\Gamma}^{1/2}
\delta
\hat{\Gamma}^{-1/2} A^{\pm})({\bf  x}) \crr -  i(\hat{\Gamma}^{-1/2}
\delta\hat{\cal   R}
\hat{\Gamma}^{-1/2} A^+)  ({\bf  x})  +  i  (\hat{\Gamma}^{-1/2}
\delta\hat{\cal R}^*
\hat{\Gamma}^{-1/2} A^-)({\bf x}).
\eeb
Therefore, formula   \r{b15} is correct up to an additive constant.
To find it, note that
$$
\delta V  |0>  =  [\frac{i}{2}  \int  d{\bf  x} d{\bf y} \phi({\bf x})
\delta \tilde{\cal R}({\bf x},{\bf y}) \phi({\bf y}) + \delta ln c] V|0>.
$$
This relation and formula \r{6*} imply
$$
<0|V^{-1} \delta V|0> = \frac{i}{2} Tr(\delta \hat{\cal  R}
\hat{\Gamma}^{-1})  +
\delta ln c.
$$
It follows from the normalization conditions
\r{b13}  and  \r{b14} that $c=(det\hat{\Gamma})^{1/4}$.
Therefore, $\delta ln c = \frac{1}{4} Tr \delta \hat{\Gamma}
\hat{\Gamma}^{-1}$. Thus,   $<0|V^{-1}\delta  V|0>  =  \frac{i}{4}  Tr
\delta(\hat{\cal R}+\hat{\cal R}^*)\Gamma^{-1}$. Formula \r{b15} is checked.

It follows  from  formula  \r{b0}  that the generators $H(A:X)$ in the
Fock representation are related with $\tilde{H}(A:X)$ by the following
relation:
$$
\breve{H}(A:X) =  H(A:X)  -  i\delta[A]  =  V_X^{-1} (\tilde{H}[A:X] -
i\delta[A]) V_X.
$$
We see that commutation relations \r{alg1} are invariant under  change
of representation.

An explicit form of operators $H(A:X)$ will be simplified if we
consider  the  case  when   the
quadratic form  $\cal  R$  is invariant under spatial translations and
rotations:
\be
\tilde{\cal R} ({\bf x},{\bf y}: u_{({\bf a},L)} X) =
\tilde{\cal R} (L^{-1}({\bf
x}- {\bf a}), L^{-1}({\bf y}-{\bf a}):X).
\l{f6}
\ee
This property implies that
$$
[\partial_k; \hat{\cal R}] = \delta_P^k \hat{\cal R};
\qquad
[\partial_k; \hat{\Gamma}^{1/2}] = \delta_P^k \hat{\Gamma}^{1/2};
$$
\be
[(x^k\partial_l - x^l\partial_k); \hat{\cal R}] = \delta_M^{kl}
\hat{\cal R};
\qquad
[(x^k\partial_l -  x^l\partial_k);   \hat{\Gamma}^{1/2}]   =
\delta_M^{kl} \hat{\Gamma}^{1/2}.
\l{f7}
\ee

The generators $H(A:X)$ are presented in table 3.

\begin{table}
\caption{Semiclassical Poincare    transformations    in    Fock
representation}
\begin{tabular}{|p{4cm}|p{14cm}|}
\hline
\parbox{4cm}{
Element of Poincare group
$(a_{\tau},\Lambda_{\tau})$
}
&
\parbox{14cm}{
Semiclassical operator $U_{a_{\tau},\Lambda_{\tau}}
(u_{a_{\tau},\Lambda_{\tau}} X \gets X) : \Psi_0 \mapsto \Psi_t$
in the Fock representation takes the initial condition  for  the
Cauchy problem to the solution of the Cauchy problem for the equation:
}
\\
\hline
\parbox{4cm}{
$a_{\tau}=0$, 
\\ 
$\Lambda_{\tau}           =          \exp(\frac{\tau}{2}
l^{sm}\theta_{sm})$; \\
spatial rotation
}
&
\parbox{14cm}{
$$
i \dot{\Psi}^{\tau} = - \frac{1}{2} \theta_{sm} {M}^{sm} \Psi_{\tau};
$$
\beb
{M}^{kl} = -  i  A^+  (x^k\partial_l  -  x^l\partial_k)  A^-
\eeb
}
\\
\hline
\parbox{4cm}{
$a_{\tau}^0 = 0$, $\Lambda_{\tau} =1$,\\
${\bf a}_{\tau} = {\bf b}\tau$;\\
spatial translation
}
&
\parbox{14cm}{
$$
i\dot{\Psi}^{\tau} = b^k {P}^k \Psi_{\tau};
$$
\beb
P^k = -iA^+ \partial_k A^-
\eeb
}
\\
\hline
\parbox{4cm}{
$a^0 = - \tau$, ${\bf a}=0$;\\
$\Lambda = 1$;\\
evolution
}
&
\parbox{14cm}{
$$
i\dot{\Psi}^{\tau} = {H}(X_{\tau}) \Psi_{\tau};
$$
\beb
{H}(X) =  \frac{1}{2} A^-{\cal H}^{--}(X)A^- + A^+
(\hat{\omega} + {\cal H}(X))A^- + \frac{1}{2} A^+ {\cal H}^{++}(X) A^+ +
\overline{H};\\
{\cal H}^{++}(X) =
\hat{\Gamma}^{-1/2} [\delta_H \hat{\cal R} - \hat{\cal R} \hat{\cal R}
- (-\Delta  +  m^2  +  V''(\Phi({\bf   x}))]   \hat{\Gamma}^{-1/2};
\\
{\cal H}^{--}(X) = ({\cal H}^{++})^+;
\\
{\cal H}(X) =
\hat{\Gamma}^{-1/2} (\hat{\cal  R}\hat{\cal R}^*
+ (-\Delta + m^2 + V''(\Phi({\bf
x})) - \frac{1}{2} \delta_H(\hat{\cal  R}  +  \hat{\cal  R}^*)
\\
+  \frac{i}{2}
[\delta_H \hat{\Gamma}^{1/2}; \hat{\Gamma}^{1/2}])\hat{\Gamma}^{-1/2}
- \hat{\omega};
\\
\hat{\omega} = \sqrt{-\Delta +m^2};
\\
{\rm formally} \qquad
\overline{H} = \overline{H}_{reg} + \frac{1}{4} Tr \hat{\Gamma};\\
\overline{H}_{reg} = - \frac{1}{4} Tr [{\cal H}^{++} + {\cal H}^{--}].
\eeb
}
\\
\hline
\parbox{4cm}{
$a_{\tau} =0$; \\
$\Lambda_{\tau} = \exp[-\tau n^kl^{0k}]$;\\
boost
}
&
\parbox{14cm}{
$$
i\dot{\Psi}^{\tau} = n^m {B}^m(X_{\tau}) \Psi_{\tau};
$$
\beb
{B}^k(X) = \frac{1}{2} A^-{\cal B}^{k --}(X)A^- + A^+
(L_k + {\cal B}^k(X))A^- + \frac{1}{2} A^+ {\cal B}^{k ++}(X) A^+ +
\overline{B^k};\\
{\cal B}^{k++}(X) =
\hat{\Gamma}^{-1/2} [\delta_k^B \hat{\cal R} - \hat{\cal R}
x^k \hat{\cal R} - (-\partial_ix^k \partial_i + x^km^2 + x^k V''(\Phi({\bf
x})))]\hat{\Gamma}^{-1/2};
\\
{\cal B}^{k--} = ({\cal B}^{k++})^+;
\\
{\cal B}^k =
\hat{\Gamma}^{-1/2} [\hat{\cal R} x^k \hat{\cal R}^* +
(-\partial_ix^k \partial_i + x^km^2 + x^k V''(\Phi({\bf x})))
- \frac{1}{2}   \delta^k_B(\hat{\cal   R}  +  \hat{\cal  R}^*)  + \crr
\frac{i}{2}
[\delta_k^B \hat{\Gamma}^{1/2}, \hat{\Gamma}^{1/2}]
] \hat{\Gamma}^{-1/2} - L_k;\\
L_k = \frac{1}{2} \hat{\omega}^{-1/2} [\hat{\omega} x^k \hat{\omega} +
(-\partial_i x^k \partial_i + x^k m^2) ] \hat{\omega}^{-1/2};\\
{\rm formally} \qquad
\overline{B^k} =   \overline{B^k}_{reg}   +   \frac{1}{4}    Tr    x^k
\hat{\Gamma};\\
\overline{B^k}_{reg} = - \frac{1}{4} Tr [{\cal B}^{k++}
+ {\cal B}^{k--}].
\eeb
}
\\
\hline
\end{tabular}
\end{table}

We see that renormalization is necessary since the evolution and boost
generators contain divergent terms
$\frac{1}{4}Tr \hat{\Gamma}$
and
$\frac{1}{4}Tr x^k\hat{\Gamma}$
which are to be changed by finite renormalizaed terms
$\frac{1}{4}Tr_R \hat{\Gamma}$
and
$\frac{1}{4}Tr_R x^k\hat{\Gamma}$.

Let us check the commutation relations between $\breve{H}(A:X)$.
Since the   divergences   arise   in   terms   $\overline{B^k}$    and
$\overline{H}$ only,  so that we suppose them to be arbitrary and then
find the conditions that provide Poincare invariance.

Let
$$
\breve{H}_k = \frac{1}{2} A^+ {\cal H}_k^{++} A^+ + A^+ {\cal  H}^{+-}
A^- + \frac{1}{2} A^-{\cal H}^{--} A^- + \overline{H_k} + i\delta_k
$$
be arbitrary quadratic Hamiltonians.  Then the property $[\breve{H}_1,
\breve{H}_2] = i\breve{H}_3$ under condition $[i\delta_1,  i\delta_2] =
i^2\delta_3$  means that
\be
{\cal H}_3^{++}  =  -  i  [
{\cal  H}_1^{+-}  {\cal  H}_2^{++}  +
{\cal H}_2^{++} ({\cal H}_1^{+-})^* -
{\cal H}_1^{++} ({\cal H}_2^{+-})^* -
{\cal  H}_2^{+-}  {\cal  H}_1^{++}]
+ \delta_1 {\cal H}_2^{++} - \delta_2 {\cal H}_1^{++}.
\l{g1}
\ee
\be
{\cal H}_3^{+-} = -i\{
{\cal H}_2^{++} ({\cal H}_1^{++})^*
- {\cal H}_1^{++} ({\cal H}_2^{++})^*
+ [{\cal H}_1^{+-}; ({\cal H}_2^{+-})]\}
+ \delta_1 {\cal H}_2^{+-} - \delta_2 {\cal H}_1^{+-},
\l{g2}
\ee
\be
\overline{H_3} = - \frac{i}{2} Tr [
{\cal H}_2^{++} ({\cal H}_1^{++})^* -
{\cal H}_1^{++}  ({\cal  H}_2^{++})^*]  +  \delta_1  \overline{H_2} -
\delta_2 \overline{H_1}.
\l{g3}
\ee
Relations \r{g1},  \r{g2},  \r{g3} are treated in  sense  of  bilinear
forms on $D(T)$.

Consider now the commutation relations.

{\bf 1.} The relations
$$
[\breve{P}^k, \breve{P}^l] = 0, \qquad [\breve{M}^{lm}, \breve{P}^s] =
i(g^{ms}\breve{P}^l - g^{ls} \breve{P}^m]
$$
are satisfied automatically since
$$
[\partial_k, \partial_l] = 0, \qquad
- [x^l\partial_m - x^m\partial_l ,  \partial_s ] = g^{ms} \partial_l -
g^{ls}\partial_m.
$$

{\bf 2.} The relation
$$
[\breve{M}^{lm}, \breve{M}^{rs}]  = -i (g^{lr} \breve{M}^{ms} - g^{mr}
\breve{M}^{ls} + g^{ms} \breve{M}^{lr} - g^{ls} \breve{M}^{mr} )
$$
is also satisfied.

{\bf 3.} For the relation
$$
[\breve{P}^k, \breve{P}^0] = 0
$$
eqs \r{g1}- \r{g3} takes the form
\be
\delta_P^k {\cal H}^{++} - [\partial_k; {\cal H}^{++}] =0,
\qquad
\delta_P^k {\cal H}^{+-} - [\partial_k; {\cal H}^{+-}] =0,
\l{g4}
\ee
\be
\delta_P^k \overline{H} =0.
\l{g4a}
\ee

4. For the relation
$$
[\breve{M}^{kl}, \breve{P}^0] =0,
$$
eqs.\r{g1} - \r{g3} are written as
\be
\delta_M^{kl} {\cal H}^{++} -
[ x^k \partial_l - x^l \partial_k ; {\cal H}^{++} ] = 0;
\qquad
\delta_M^{kl} {\cal H}^{+-} -
[ x^k \partial_l - x^l \partial_k ; {\cal H}^{+-} ] = 0;
\l{g5}
\ee
\be
\delta_M^{kl} \overline{H} = 0.
\l{g5a}
\ee

5. Consider the relation
$$
[\breve{M}^{k0}, \breve{P}^s] = -ig^{ks}\breve{P}^0.
$$
We write eqs.\r{g1} - \r{g3} as follows:
\be
[\partial_s ,  {\cal B}^{k++} ] - \delta_P^s {\cal B}^{k++} = - g^{ks}
{\cal H}^{++},
\qquad
[\partial_s ,  {\cal B}^{k+-} ] - \delta_P^s {\cal B}^{k+-} = - g^{ks}
{\cal H}^{+-},
\l{g6}
\ee
\be
\delta_P^s \overline{B^k} = g^{ks} \overline{H}.
\l{g6a}
\ee

6. The commutation relation
$$
[\breve{M}^{lm}, \breve{M}^{k0} ] = -i (g^{lk} \breve{M}^{m0} - g^{mk}
\breve{M}^{l0})
$$
is equivalent to
\beb
[x^l\partial_m -  x^m  \partial_l  ;  {\cal B}^{k++} ] - \delta_M^{lm}
{\cal B}^{k++} = g^{lk} {\cal B}^{m++} - g^{mk} {\cal B}^{l++},
\eeb
\bea
[x^l\partial_m -  x^m  \partial_l  ;  {\cal B}^{k+-} ] - \delta_M^{lm}
{\cal B}^{k+-} = g^{lk} {\cal B}^{m+-} - g^{mk} {\cal B}^{l+-},
\l{g7}
\eea
\be
- \delta_M^{kl}   \overline{B^k}  =  g^{lk}  \overline{B^m}  -  g^{mk}
\overline{B^l}.
\l{g7a}
\ee

7. The most nontrivial commutation relations are
$$
[\breve{M}^{k0}; \breve{P}^0] = i \breve{P}^k,
\qquad
[\breve{M}^{k0}; \breve{M}^{l0}] = -i\breve{M}^{kl}
$$
They can be rewritten as follows:
\bea
0 = -i \{
{\cal B}^{k+-} {\cal H}^{++} +
{\cal H}^{++} ({\cal B}^{k+-})^*
-
{\cal B}^{k++} ({\cal H}^{+-})^* -
{\cal H}^{+-} ({\cal B}^{k++})
\} + \delta_B^k {\cal H}^{++} - \delta_H {\cal B}^{k++};
\crr
-i \partial_k = -i \{
{\cal H}^{++} {\cal B}^{k++}
- {\cal B}^{k++} ({\cal H}^{++})^* +
[{\cal B}^{k+-}; {\cal H}^{+-}]
\} + \delta_B^k {\cal H}^{+-} - \delta_H {\cal B}^{k+-};
\l{g8}
\eea
\be
0 = -  \frac{i}{2}  Tr  [{\cal  H}^{++}  ({\cal  B}^{k++})^*  -  {\cal
B}^{k++} ({\cal  H}^{++})^*]  +  \delta_B^k  \overline{H}  -  \delta_H
\overline{B^k}
\l{g8a}
\ee
and
\bea
0 = -i \{
{\cal B}^{k+-} {\cal B}^{l++} +
{\cal B}^{l++} ({\cal B}^{k+-})^*
-
{\cal B}^{k++} ({\cal B}^{l+-})^* -
{\cal B}^{l+-} ({\cal B}^{k++})
\} + \delta_B^k {\cal B}^{l++} - \delta_B^l {\cal B}^{k++};
\crr
i(x^k \partial_l - x^l\partial_k) = -i \{
{\cal B}^{l++} {\cal B}^{k++}
- {\cal B}^{k++} ({\cal B}^{l++})^* +
[{\cal B}^{k+-}; {\cal B}^{l+-}]
\} + \delta_B^k {\cal B}^{l+-} - \delta_B^l {\cal B}^{k+-};
\l{g9}
\eea
\be
0 = -  \frac{i}{2}  Tr  [{\cal  B}^{l++}  ({\cal  B}^{k++})^*  -  {\cal
B}^{k++} ({\cal B}^{l++})^*] + \delta_B^k \overline{B^l} - \delta_B^l
\overline{B^k}.
\l{g9a}
\ee

{\bf 3.}  Properties \r{g4},  \r{g5},  \r{g6},  \r{g7} are obvious
corollaries of relations \r{f7}.  Properties  \r{g8}  and  \r{g9}  are
checked by nontrivial but also direct computations.

Properties \r{g4a}, \r{g5a}, \r{g6a}, \r{g7a}, \r{g8a},
\r{g9a} will  be  satisfied  if  the renormalized trace satisifies the
following properties:
\bea
\delta^P_k Tr_R \hat{\Gamma} = 0; \qquad
\delta^M_{kl} Tr_R\hat{\Gamma} = 0;
\crr
\delta^P_l Tr_R x^k \hat{\Gamma} = - \delta^{kl} Tr_R \hat{\Gamma};
\qquad
\delta^M_{kl} Tr_R x^k\hat{\Gamma} = \delta^{kl} Tr_R x^m \hat{\Gamma}
- \delta^{mk} Tr_R x^l \hat{\Gamma};\crr
Tr [x^l  (\delta^B_k  \hat{\Gamma}  -  \hat{\cal  A}  x^k  \hat{\Gamma}  -
\hat{\Gamma} x^k \hat{\cal A})
- x^k  (\delta^B_l  \hat{\Gamma}  -  \hat{\cal  A}  x^l \hat{\Gamma} -
\hat{\Gamma} x^l \hat{\cal A})]
+ \delta_l^B Tr_R x^k\hat{\Gamma} - \delta_k^B Tr_R x^l \hat{\Gamma} =
0; \crr
Tr [x^l  (\delta^H  \hat{\Gamma}  -  \hat{\cal   A}   \hat{\Gamma}   -
\hat{\Gamma} \hat{\cal A})
- (\delta^B_l  \hat{\Gamma}  -  \hat{\cal  A}   x^l   \hat{\Gamma}   -
\hat{\Gamma} x^l \hat{\cal A})]
+ \delta_l^B Tr_R \hat{\Gamma} - \delta^H Tr_R x^l \hat{\Gamma} = 0,
\l{trace}
\eea
where $\hat{\cal A}  = \frac{1}{2} (\hat{\cal R} + \hat{\cal R}^*)$.

Thus, algebraic commutation relations are checked.

\subsection{Conditions of integrability}

The problem of reconstructing a representation of a  local  Lie  group
from a  representation  of  a Lie algebra ("integrability problem") is
mathematically nontrivial.  Different conditions of integrability were
presented in \c{BR,N,FS1,FS2,FS3}.

The problem  of  reconstructing  the operators $U_g(u_gX \gets X)$ and
checking the group property was discussed in details  in  \c{Shvedov}.
It has been shown that the operators $U_g(u_gX \gets X)$ are correctly
defined under the following sufficient conditions.

Let $h(\alpha)$ be an arbitrary smooth curve on the Poincare group.

P1, {\it
For self-adjoint operators
$$
A_k = L_k,  \qquad
A_{d+k} = - i\partial_k, \qquad
A_{2d+kd+l} = - i(x^k\partial_l - x^l\partial_k),
\qquad A_{2d+d^2+1} = \hat{\omega}
$$
there exists such a positively definite operator $T$ that\\
1. $||T^{-1/2}A_jT^{-1/2}|| <\infty$, $||A_jT^{-1}|| < \infty$.\crr
2. for all $t_1$ there exists such a constant $C$ that
$||T^{1/2} e^{iA_jt} T^{-1/2}|| \le C$,  $||Te^{-iA_jt} T^{-1}|| \le
C$, $t \in [-t_1,t_1]$.
}

P2. {\it The  $\alpha$-dependent  operator  functions  $T  {\cal   B}^{k++}
(u_{h(\alpha)}X)$ and $T{\cal H}^{++}(u_{h(\alpha)}X)$ are continuous in
the Hilbert-Schmidt topology $||\cdot||_2$.}

P3. {\it The     $\alpha$-dependent     operator      functions      ${\cal
B}^{k++}(u_{h(\alpha)}X)$ and   ${\cal   H}^{++}(u_{h(\alpha)}X)$  are
continuously differentiable   with   respect   to   $\alpha$   in   the
Hilbert-Schmidt topology.}

P4. {\it The      $\alpha$-dependent      operator     functions     ${\cal
B}^k(u_{h(\alpha)}X)$, ${\cal       H}(u_{h(\alpha)}X)$,
$T{\cal B}^k(u_{h(\alpha)}X)T^{-1}$,
$T^{1/2}{\cal B}^k(u_{h(\alpha)}X)T^{-1/2}$,
$T{\cal H}(u_{h(\alpha)}X)T^{-1}$,
$T^{1/2}{\cal H}(u_{h(\alpha)}X)T^{-1/2}$ are strongly continuous.}

P5. {\it The  $\alpha$-dependent   operator   functions
$T^{-1/2}   {\cal H}(u_{h(\alpha)}X)T^{-1/2}$,
$T^{-1/2}   {\cal B}^k(u_{h(\alpha)}X)T^{-1/2}$,
${\cal H}(u_{h(\alpha)}X)T^{-1}$,
${\cal B}^k(u_{h(\alpha)}X)T^{-1}$
are continuously   differentiable  with  respect  to  $\alpha$  in  the
operator norm $||\cdot||$ topology.}

P6. {\it The functions $\overline{H}(u_{h(\alpha)}X)$ and
 $\overline{B^k}(u_{h(\alpha)}X)$ are continuous.}

The property P6 can be substituted by the following property.

P6'.
{\it
(a) The operators ${\cal B}^{k++}$ and  ${\cal  H}^{++}$  are  of  the
trace class and $Tr {\cal B}^{k++}(u_{h(\alpha)}X)$
and $Tr {\cal  H}^{++}(u_{h(\alpha)}X)$  are  continuous  functions  of
$\alpha$. \\
(b) The functions $Tr_R \Gamma(u_{h(\alpha)}X)$ and
$Tr_R x^k\Gamma(u_{h(\alpha)}X)$ are continuous.
}

Let us first justify property P1.

Let
$$
\hat{K} = \hat{\omega}^{-1/4} ({\bf x}^2 + 1)^{-1} \hat{\omega}^{-1/4}.
$$
This is a bounded self-adjoint positively definite operator without zero
eigenvalues. Therefore, $\hat{K}^{-1} \equiv T^{1/2}$ is a (non-bounded)
self-adjoint operator and
$$
T= \hat{\omega}^{1/4}  ({\bf x}^2 + 1) 
\hat{\omega}^{1/2}
({\bf x}^2 + 1) \hat{\omega}^{1/4};
$$
$T \ge c>0$ for some $c$.

The  first  part  of property P1 is justified as follows.  One
should check that the following norms are finite:
\beb
||\hat{\omega}^{-1/4} ({\bf x}^2+1)^{-1} \hat{\omega}^{-1/4}
\hat{\omega} \hat{\omega}^{-1/4}          ({\bf           x}^2+1)^{-1}
\hat{\omega}^{-1/4} ||; \crr
||\hat{\omega}^{-1/4} ({\bf x}^2+1)^{-1} \hat{\omega}^{-1/4}
\hat{\omega} x^s \hat{\omega}^{-1/4}          ({\bf           x}^2+1)^{-1}
\hat{\omega}^{-1/4} ||; \crr
||\hat{\omega}^{-1/4} ({\bf x}^2+1)^{-1} \hat{\omega}^{-1/4}
(\hat{k}^j x^s - \hat{k}^s x^j)
\hat{\omega}^{-1/4}          ({\bf           x}^2+1)^{-1}
\hat{\omega}^{-1/4} ||; \crr
||\hat{\omega}^{-1/4} ({\bf x}^2+1)^{-1} \hat{\omega}^{-1/4}
\hat{k}^j \hat{\omega}^{-1/4}          ({\bf           x}^2+1)^{-1}
\hat{\omega}^{-1/4} ||; \crr
||\hat{\omega}\hat{\omega}^{-1/4} ({\bf x}^2+1)^{-1} \hat{\omega}^{-1/4}
 \hat{\omega}^{-1/4}          ({\bf           x}^2+1)^{-1}
\hat{\omega}^{-1/4} ||; \crr
||\hat{\omega} x^s \hat{\omega}^{-1/4} ({\bf x}^2+1)^{-1} \hat{\omega}^{-1/4}
 \hat{\omega}^{-1/4}          ({\bf           x}^2+1)^{-1}
\hat{\omega}^{-1/4} ||; \crr
||
(\hat{k}^j x^s - \hat{k}^s x^j)
\hat{\omega}^{-1/4} ({\bf x}^2+1)^{-1} \hat{\omega}^{-1/4}
\hat{\omega}^{-1/4}          ({\bf           x}^2+1)^{-1}
\hat{\omega}^{-1/4} ||; \crr
||\hat{k}^j
\hat{\omega}^{-1/4} ({\bf x}^2+1)^{-1} \hat{\omega}^{-1/4}
\hat{\omega}^{-1/4}          ({\bf           x}^2+1)^{-1}
\hat{\omega}^{-1/4} ||,
\eeb
where $\hat{k}^j = - i \partial/\partial x^j$.

To check this statement, it is sufficient to notice that
lemma A.29 of Appendix A implies that the operators 
\be
[\hat{\omega}^{\alpha}; ({\bf x}^2+1)^{-1} ];
\qquad
[\hat{\omega}^{\alpha}; x^s({\bf x}^2+1)^{-1} ];
\qquad
[\hat{\omega}^{\alpha}; x^lx^s({\bf x}^2+1)^{-1} ]
\l{l3.6}
\ee
are bounded if $\alpha \le 1$.

To prove the second part of P1,  represent  it  in  the
following form:
\be
||e^{iA_jt}T^{1/2}e^{-iA_jt} T^{-1/2}||    \equiv    ||   T_j^{1/2}(t)
T^{-1/2}|| \le C; \qquad ||T_j(t)T^{-1}|| \le C.
\l{f11}
\ee
It is necessary to investigate the Poincare transformation  properties
of the operators $\hat{x}^j$ and $\hat{k}^j$.

Notice that the following relations are satisfied:
\beb
e^{i\hat{\omega}t} \hat{x}^l   e^{-i\hat{\omega}t}   =   \hat{x}^l   +
\hat{k}^l \hat{\omega}^{-1} t,
\qquad
e^{i\hat{\omega}t} \hat{k}^l   e^{-i\hat{\omega}t}   =   \hat{k}^l;
\crr
e^{i\hat{k}^sa^s} \hat{x}^l  e^{-  i\hat{k}^sa^s}  =  \hat{x}^l + a^l;
\qquad
e^{i\hat{k}^sa^s} \hat{k}^l  e^{-  i\hat{k}^sa^s}  =  \hat{k}^l;
\crr
e^{\frac{i\tau}{2} \theta_{ms}    (\hat{x}^m\hat{k}^s    -   \hat{x}^s
\hat{k}^m)}
\hat{x}^l
e^{- \frac{i\tau}{2} \theta_{ms}    (\hat{x}^m\hat{k}^s    -   \hat{x}^s
\hat{k}^m)} =
(e^{-\tau\theta} \hat{x})^l;
\crr
e^{\frac{i\tau}{2} \theta_{ms}    (\hat{x}^m\hat{k}^s    -   \hat{x}^s
\hat{k}^m)}
\hat{k}^l
e^{- \frac{i\tau}{2} \theta_{ms}    (\hat{x}^m\hat{k}^s    -   \hat{x}^s
\hat{k}^m)} =
(e^{-\tau\theta} \hat{k})^l;
\crr
e^{iL^1\tau} \hat{k}^l e^{-iL^1\tau} = \hat{k}^l , \quad  l \ge 2;
\qquad
e^{iL^1\tau} \hat{k}^1   e^{-iL^1\tau}   =   \hat{k}^1   \cosh   \tau   -
\hat{\omega} \sinh\tau.
\eeb
The operators $\hat{X}^l(\tau) = e^{iL^1\tau} \hat{x}^l e^{-iL^1\tau}$
have the following Weyl symbols:
$$
X^1 = \frac{\omega_{\bf k}}{\omega_{\bf  k}\cosh\tau  -  k^1\sinh\tau}  x^1;
\qquad
X^{\alpha} = x^{\alpha} + \frac{k^{\alpha}\sinh\tau x^1}
{\omega_{\bf  k}\cosh\tau  -  k^1\sinh\tau}
$$

To check  the  properties,  it  is  sufficient  to  show that they are
satisfied at $\tau=0$ and show that the derivatives of  left-hand  and
right-hand sides of these relations coincide.

Making use    of   commutation   relations   $[x^s,f(\hat{k})]   =   i
\frac{\partial f }{\partial k^s}(\hat{k})$ and boundedness
of the operators \r{l3.6},  we
find that  operators  \r{f11} are bounded uniformly with respect to $t
\in [0,t_1]$. property P1 is checked.

\subsection{Choice of the operator $\cal R$}

Let us  choose operator $\cal R$ in order to satisfy properties P1-P5,
P7. We will use the notions of Appendix A (subsection A.5).  First, we
construct such    an   asymptotic   expansion   of   a   Weyl   symbol
$\underline{\cal R}_N$ that for $\underline{\cal R} =  \underline{\cal
R}_N$
\bea
deg [\delta_B^l \underline{\cal  R}  -  \underline{\cal  R}  *  x^l  *
\underline{\cal R}   -   x^l  (\omega_k^2  +  V''(\Phi({\bf  x})))]  >
\max\{d/2, d-1\}; \crr
deg [\delta_H \underline{\cal  R}  -  \underline{\cal  R}  *
\underline{\cal R}   -    (\omega_k^2  +  V''(\Phi({\bf  x})))]  >
\max\{d/2, d-1\}.
\l{p1}
\eea
Next, we  will construct another asymptotic expansion of a Weyl symbol
$\underline{\cal R}$ which obeys the condition $Im \underline{\cal  R}
> 0$  and  approximately  equals  to  $\underline{\cal  R}_N$ at large
$|k|$, so that eqs.\r{p1} are satisfied.

This will imply that properties P2-P5, P6' are satisfied.

Let us define the expansions $\underline{\cal R}_N$ with the  help  of
the following recursive relations. Set
\bea
\underline{\cal R}_0 = i \omega_k; \crr
\underline{\cal S}_n    =    -   \delta_H   \underline{\cal   R}_n   +
\underline{\cal R}_n   *   \underline{\cal   R}_n   +   \omega_k^2   +
V''(\Phi({\bf x}));\crr
\underline{\cal R}_{n+1} = \underline{\cal R}_n +  \frac{i}{2\omega_k}
\underline{\cal S}_{n}.
\l{p1a}
\eea

{\bf Lemma 3.1.} {\it The following relation is satisfied:
$$
deg \underline{\cal S}_n = n.
$$
}

{\bf Proof.}  For $n=0$,  $\underline{\cal S}_0 = V''(\Phi({\bf x}))$,
so that statement of lemma is satisfied.  Suppose  that  statement  of
lemma is justified for $n<N$. Check it for $n=N$. One has
$$
\underline{\cal S}_N = \underline{\cal S}_{N-1} + \underline{\cal R}_N
* \left( \frac{i}{2\omega_k} \underline{\cal S}_{N-1} \right)
+ \left( \frac{i}{2\omega_k} \underline{\cal S}_{N-1} \right)
* \underline{\cal R}_N +
\left( \frac{i}{2\omega_k} \underline{\cal S}_{N-1} \right)
* \left( \frac{i}{2\omega_k} \underline{\cal S}_{N-1} \right)
- \frac{i}{2\omega_k} \delta_H \underline{\cal S}_{N-1}.
$$
Since
$$
deg\left[\left( \frac{i}{2\omega_k} \underline{\cal S}_{N-1} \right)
* \left( \frac{i}{2\omega_k} \underline{\cal S}_{N-1} \right)
- \frac{i}{2\omega_k} \delta_H \underline{\cal S}_{N-1}\right] \ge deg
\underline{\cal S}_{N-1} + 1 = N
$$
and
\beb
\underline{\cal S}_N = \underline{\cal S}_{N-1} + \underline{\cal R}_N
* \left( \frac{i}{2\omega_k} \underline{\cal S}_{N-1} \right)
+ \left( \frac{i}{2\omega_k} \underline{\cal S}_{N-1} \right)
* \underline{\cal R}_N \simeq
\crr
\underline{\cal S}_{N-1} + \underline{\cal R}_N
 \left( \frac{i}{2\omega_k} \underline{\cal S}_{N-1} \right)
+ \left( \frac{i}{2\omega_k} \underline{\cal S}_{N-1} \right)
 \underline{\cal R}_N = 0
\eeb
up to terms of the degree $N$, one finds
$$
deg \underline{\cal S}_N = N.
$$
Lemma 3.1 is proved.

Denote
$$
\underline{X}_n^l =    -    \delta_B^l    \underline{\cal    R}_n    +
\underline{\cal R}_n * x^l * \underline{\cal R}_n + x^l (\omega_k^2  +
V''(\Phi({\bf x}))).
$$

{\bf Lemma 3.2.} {\it The following property is obeyed:
\be
\delta_B^l \underline{\cal S}_n - \delta_H \underline{X}_n^l = -
\underline{X}_n^l * \underline{\cal R}_n -
\underline{\cal R}_n * \underline{X}_n^l + \underline{\cal S}_n *  x^l
* \underline{\cal  R}_n + \underline{\cal R}_n * x^l * \underline{\cal
S}_n.
\l{p2}
\ee
}

{\bf Proof.} Denote
$$
\underline{F}_n^l =
\delta_B^l \underline{\cal S}_n - \delta_H \underline{X}_n^l
+ \underline{X}_n^l * \underline{\cal R}_n +
\underline{\cal R}_n * \underline{X}_n^l - \underline{\cal S}_n *  x^l
* \underline{\cal  R}_n - \underline{\cal R}_n * x^l * \underline{\cal
S}_n.
$$
One has
\beb
\underline{F}_n^l = (\delta_B^l - x^l \delta_H) V''(\Phi({\bf  x}))  +
[\delta_H; \delta_B^l]  \underline{\cal  R}_n  -
[x^l  (\omega_k^2  + \crr
V''(\Phi({\bf x})) - (\omega_k^2  + V''(\Phi({\bf x})) * x^l ] *
\underline{\cal R}_n +
\underline{\cal R}_n *
[x^l  (\omega_k^2  +
V''(\Phi({\bf x})) - x^l * (\omega_k^2  + V''(\Phi({\bf x}))]
\eeb
It follows from the definition of the Weyl symbol that
$$
x^l * f(x,k) = (x^l + \frac{i}{2} \frac{\partial}{\partial k^l}) f(x,k)
$$
One also has
$$
(\delta_B^l - x^l \delta_H) V''(\Phi({\bf  x})) = 0.
$$
Thus,
$$
\underline{F}_n^l = [\delta_H; \delta_B^l] \underline{\cal R}_n +
ik^l * \underline{\cal R}_n - \underline{\cal R}_n * ik^l =
\frac{\partial \underline{\cal   R}_n}{\partial  x^l  }  -  \delta_P^l
\underline{\cal R}_n.
$$
However, the property
$$
\frac{\partial \underline{\cal  R}_n}{\partial   x^l}   =   \delta_P^l
\underline{\cal R}_n
$$
which means that eq.\r{f6} is satisfied is checked by induction. Lemma
3.2 is proved.

{\bf Lemma 3.3.} {\it The following properties are satisfied:\\
1. $deg \underline{X}_n^l = n$. \\
2. $deg(\underline{X}_n^l - x^l \underline{\cal S}_n) \ge n+1$.
}

{\bf Proof.} It follows from the results of Appendix A that
$\underline{X}_n^l$ is an asymptotic expansion of a Weyl symbol. Let
$ deg \underline{X}_n^l = \alpha$.

Suppose that $\alpha <n$.  Then the left-hand side of eqs.\r{p2} is of
the degree $\alpha$, the degree of the right-hand side of eq.\r{p2}
is greater than or equal to
$\alpha-1$. In the leading order in $1/|k|$ the  right-hand  side  has
the form
one has $(-2i\omega_k
\underline{X}_n^l)$ and  its degree should be greater than or equal to
$\alpha$ .  Therefore, $deg \underline{X}_n^l \ge
\alpha + 1$. We obtain a contradiction.

Suppose $\alpha>n$.  Then  the  left-hand  side of eq.\r{p2} is of the
degree $n$,  the right-hand side in the leading order in  $1/|k|$  has
the form $2i\omega_k x^l \underline{\cal S}_n$. so that $deg
\underline{\cal S}_n$ should obey the inequality
$deg \underline{\cal S}_n \ge n+1$. We also obtain a contradiction.

Thus, $\alpha=n$. In the leading order in $1/|k|$ one has
$$
0 \simeq -2i\omega_k (\underline{X}_n^l - x^l \underline{\cal S}_n)
$$
up to terms of the degree $n$, so that
$deg (\underline{X}_n^l - x^l \underline{\cal S}_n)  \ge  n+1$.  Lemma
3.3 is proved.

We see  that  for  $N\ge  max\{d/2,d-1\}$  the  properties  \r{p1} are
satisfied.

{\bf Lemma 3.4.} {\it Let $\underline{\cal R}^{(1)}$ and
$\underline{\cal R}^{(2)}$  be  asymptotic expansions of Weyl symbols,
$deg \underline{\cal R}^{(1)} = deg \underline{\cal R}^{(2)} = -1$ and
$deg(\underline{\cal R}^{(1)} - \underline{\cal R}^{(2)}) = N+1$. Then
$$
deg (\underline{X}^{(1)l} - \underline{X}^{(2)l}) = N
$$
and
$$
deg (\underline{\cal S}^{(1)} - \underline{\cal S}^{(2)}) = N.
$$
}

{\bf Proof.}   Denote   $\underline{\cal  R}^{(1)}  -  \underline{\cal
R}^{(2)} = \underline{D}$. Then
$$
\underline{X}^{(1)l} -    \underline{X}^{(2)l}    =    -    \delta_B^l
\underline{D} +  \underline{\cal  R}^{(1)}  *  x^l  *  \underline{D} +
\underline{D} * x^l * \underline{\cal R}^{(1)} + \underline{D} *
\underline{D} * x^l * \underline{D}.
$$
We see  that  $deg(\underline{X}^{(1)l}  - \underline{X}^{(2)l}) = N$.
The second statement is checked analogously. Lemma 3.4 is proved.

Let us construct such an  asymptotic  expansion  $\underline{\cal  R}$
that $deg(\underline{\cal  R}  -  \underline{\cal R}_N) = N+1$ and $Im
\underline{\cal R} > 0$.  We will look  for  $\underline{\cal  R}$  as
follows (cf. \c{MS3}),
$$
\underline{\cal R}  =  \underline{\cal  A}  +  i \omega_k^{1/4} * \exp
\underline{\cal B}  *  \omega_k^{1/4}  *  \exp  \underline{\cal  B}  *
\omega_k^{1/4},
$$
where $\underline{\cal  A}$  and  $\underline{\cal  B}$ are {\it real}
asymptotic expansions. Then
\beb
\underline{\Gamma^{1/2}} = \omega_k^{1/4} * \exp \underline{\cal B}  *
\omega_k^{1/4}; \crr
\underline{\Gamma^{-1/2}} =  \omega_k^{-1/4}  * \exp (-\underline{\cal
B})  * \omega_k^{-1/4}
\eeb
are also asymptotic expansions of Weyl symbols. Choose
$\underline{\cal A}$ and $\underline{\cal B}$ to be polynomials,
$$
\underline{\cal A}  =   \sum_{s=1}^{S_1}   \frac{A_s({\bf   x},   {\bf
k}/\omega_{\bf k})}{\omega_{\bf k}^{2s}},
\qquad
\underline{\cal B}  =   \sum_{s=1}^{S_2}   \frac{B_s({\bf   x},   {\bf
k}/\omega_{\bf k})}{\omega_{\bf k}^{2s}},
$$
where $S_1 = [N/2]$, $S_2 = [\frac{N+1}{2}]$.

{\bf Lemma 3.5.} {\it  There  exists  unique  functions  $A_1$,  ...,
$A_{S_1}$, $B_1$, ..., $B_{S_2}$ such that
$deg (\underline{\cal R} - \underline{\cal R}_N) = N+1$.
}

{\bf Proof.} It follows from recursive relations \r{p1a} that
\beb
Re \underline{\cal  R}_N  = \sum_{s=1}^{\infty} \frac{A_{N,s}({\bf x},
{\bf k}/\omega_{\bf k})}{\omega_{\bf k}^{2s}}, \crr
Im \underline{\cal R}_N = \omega_{\bf k}  +
 \sum_{s=1}^{\infty} \frac{C_{N,s}({\bf x},
{\bf k}/\omega_{\bf k})}{\omega_{\bf k}^{2s}}.
\eeb
Therefore, $A_s  = A_{N,s}$,  so that $\underline{\cal A}$ is uniquely
defined. Denote
$$
\underline{{\cal B}_s} = \frac{B_{s}({\bf x},
{\bf k}/\omega_{\bf k})}{\omega_{\bf k}^{2s}}.
$$
Show that  $\underline{{\cal B}_s}$ is uniquely defined.  In the leading
order in $1/|{\bf k}|$, one has
$$
Im \underline{\cal R} \simeq \omega_{\bf k}  +  2\underline{{\cal  B}_1}
\omega_{\bf k},
$$
so that   $B_1   =   C_{N,1}/2$.   Suppose   that   one   can   choose
$\underline{{\cal B}_1}$,  ..., $\underline{{\cal B}_{s-1}}$ in such a way
that the degree of the asymptotic expansion of a Weyl symbol
$$
\underline{F}_{N,s} = Im \underline{\cal R}_N - \omega_{\bf k}^{1/4} *
\exp(\underline{{\cal B}_1} + ... + \underline{{\cal B}_{s-1}}) *
\omega_{\bf k}^{1/2} *
\exp(\underline{{\cal B}_1} + ... + \underline{{\cal B}_{s-1}})
* \omega_{\bf k}^{1/4}
$$
satisfies the inequality
$$
deg \underline{F}_{N,s} \ge 2s-1.
$$
Choose $\underline{{\cal B}_s}$ in such a way that
$deg \underline{F}_{N,s} \ge 2s-1$. One has
$$
\underline{F}_{N,s+1} = Im \underline{\cal R}_N - \omega_{\bf k}^{1/4}
* \sum_{l_1=0}^{\infty} \frac{( \underline{{\cal B}_1} + ...
+ \underline{{\cal B}_{s-1}}+ \underline{{\cal B}_s})^{l_1}}{l_1!}
* \omega_{\bf k}^{1/2} *
\sum_{l_2=0}^{\infty} \frac{( \underline{{\cal B}_1} + ...
+ \underline{{\cal B}_{s-1}}+ \underline{{\cal B}_s})^{l_2}}{l_2!}
* \omega_{\bf k}^{1/4}.
$$
Up to terms of the degree $2s+1$, one has
\beb
\underline{F}_{N,s+1} \simeq Im \underline{\cal R}_N - \omega_{\bf k}^{1/4}
* (\sum_{l_1=0}^{\infty} \frac{( \underline{{\cal B}_1} + ...
+ \underline{{\cal B}_{s-1}}  )^{l_1}}{l_1!}
+ \underline{{\cal B}_s})
* \omega_{\bf k}^{1/2} *
(\sum_{l_2=0}^{\infty} \frac{( \underline{{\cal B}_1} + ...
+ \underline{{\cal B}_{s-1}} )^{l_2}}{l_2!}
+ \underline{{\cal B}_s})
* \omega_{\bf k}^{1/4}
\crr
\simeq
\underline{F}_{N,s} - 2 \underline{{\cal B}_s} \omega_{\bf k}.
\eeb
Since
$$
\underline{F}_{N,s} =       \frac{1}{\omega_{\bf      k}^{2s-1}      }
\sum_{l=0}^{\infty} \frac{F_{N,s,l}  ({\bf  x},  {\bf   k}/\omega_{\bf
k})}{\omega_{\bf k}^l},
$$
one finds that
$$
\underline{{\cal B}_s} = \frac{1}{2\omega_k^{2s}}
{F_{N,s,0}  ({\bf  x},  {\bf   k}/\omega_{\bf
k})}
$$
is uniquely defined. Lemma 3.5 is proved.

Thus, we have constructed the operator $\cal R$ such that properties
\r{p1} are satisfied. We obtain the following theorem.

{\bf Theorem 3.6.} {\it Properties \r{f6}, P2-P5, P6'(a) are satisfied.}

This theorem  is  a  direct  corollary  of  the results of Appendix A.
Property P1 is satisfied because of construction of the operator $\cal
R$. Properties P2-P5, P6'(a) are corollaries of Theorems A.31, A.32, A.33,
properties \r{p1} and lemmas A.8, A.9, A.19.

\subsection{Regularization and renormalization of a trace}

The purpose of this subsection is to specify functionals  $Tr_R\Gamma$
and $Tr_R  x^k\Gamma$  of arguments $\Phi$,  $\Pi$ in order to satisfy
properties P6'(b), \r{trace}.
We want the renormalized trace to satisfy properties
like these:\\
(i) $Tr_R \hat{A} = Tr \hat{A}$ if $\hat{A}$ is of the trace class;\\
(ii) $Tr_R (\hat{A}+\lambda \hat{B})  =  Tr_R  \hat{A}  +  \lambda  Tr_R
\hat{B}$; \\
(iii) $Tr_R [\hat{A}; \hat{B}]  = 0$; \\
(iv) $Tr_R \hat{A}_n \to 0$ if $A_n \to 0$ \\
for such  class of operators that is as wide as possible.  Under these
conditions, properties P6'(b) and \r{trace}
are satisfied.  However,  one  cannot
specify such a renormalized trace. Namely, one should have
\be
Tr_R [\hat{x}_j;  {\cal W}(\frac{k_j}{\omega_{\bf k}^l} f({\bf x}))] =
0,
\l{p3a}
\ee
where $f \in S({\bf R}^d)$.
${\cal W}(A)$ is a Weyl quantization of the  function  $A$  (see
appendix A). Property \r{p3a} means that
$$
Tr_R {\cal W}(i \frac{\partial}{\partial  k_i}  \frac{k_j}{\omega_{\bf
k}^l} f({\bf x})) = 0.
$$
Therefore,
\be
\delta_{ij} Tr_R {\cal W} (\frac{f({\bf x})}{\omega_{\bf k}^l})
- l Tr_R {\cal W}(\frac{k_ik_j}{\omega_{\bf k}^{l+2}} f({\bf x})) = 0.
\l{p4}
\ee
Choose $l=d$.  Consider  $i=j$  in eq.\r{p4} and perform the summation
over $i$.  Making use of the relation $\omega_{{\bf k}}^2 -  k_ik_i  =
m^2$, we find
$$
Tr_R {\cal W}(m^2 \omega_{\bf k}^{-d-2} f({\bf x})) = 0.
$$
However, the operator with Weyl symbol  $m^2  f({\bf  x})  \omega_{\bf
k}^{-d-2}$ is of the trace class.  Its trace is nonzero, provided that
$\int d{\bf x} f(x) \ne 0$.

However, we can introduce a notion of  a  trace  for  {\it  asymptotic
expansions of  Weyl  symbols.} The trace will be specified not only by
operator but also by its asymptotic expansion which is not unique (see
remark after definition A.6).

Let $\underline{A} = (A, \check{A})$ be asymptotic expansion of a Weyl
symbol. Suppose that the coefficients $A_l$ of the  formal  asymptotic
expansion
$$
\check{A} \equiv  \sum_{l=0}^{\infty}  \omega_{\bf  k}^{-\alpha-l} A_l
(x, {\bf k}/\omega_{\bf k})
$$
are polynomial in ${\bf k}/\omega_{\bf k}$. One formally has
\be
Tr_R \underline{A} = \sum_{l=0}^{l_0}  \int \frac{d{\bf k}
d{\bf x} }{(2\pi)^d} \frac{1}{\omega_{\bf k}^{\alpha+l}}
A_l ({\bf x}, {\bf k}/\omega_{\bf k})
+ \int \frac{d{\bf k} d{\bf x}}{(2\pi)^d} (A({\bf x},{\bf k}) -
\sum_{l=0}^{l_0} \frac{1}{\omega_{\bf k}^{\alpha+l}}
A_l ({\bf x}, {\bf k}/\omega_{\bf k})).
\l{p5}
\ee
For $\alpha + l_0 + 1 > d$,  the last integral in the right-hand  side
of eq.\r{p5}  converges.  To  specify trace,  it is sufficient then to
specify values of integrals
\be
I^{s,n}_{i_1...i_n} =   \int   \frac{d{\bf    k}}{\omega_{\bf    k}^s}
\frac{k_{i_1}}{\omega_{\bf k}} ... \frac{k_{i_n}}{\omega_{\bf k}}
\l{p6}
\ee
for $s\le d$ which  are  divergent.  We  will  define  the  quantities
\r{p6}, making use of the following argumentation.

1. We are going to specify to specify trace in such a way that
\be
Tr_R \frac{\partial}{\partial k_i} \underline{A} = 0.
\l{p7}
\ee
Let
$$
\underline{A} = \frac{1}{\omega_{\bf k}^{s-1}}
\frac{k_{j_1}}{\omega_{\bf k}} ... \frac{k_{j_{n+1}}}{\omega_{\bf k}}
$$
property \r{p7} implies the following recursive relations
\be
\sum_{s=1}^{n+1} \delta_{ij_s}        I^{s,n}_{j_1...j_{s-1}j_{s+1}...
j_{n+1}} = (s+n) I^{s,n+2}_{ij_1...j_{n+1}}.
\l{p8}
\ee
Therefore, $I^{s,n}=0$  for  odd $n$,  while for even $n$ $I^{s,n}$ is
defined from eqs.\r{p8},  for example,  $I^{s,2}_{ij} =  \frac{1}{s}
\delta_{ij} I^{s,0}$. Therefore, it is sufficient to define integrals
\be
I^{s,0} = \int d{\bf k} \omega_{\bf k}^{-s}.
\l{p9}
\ee

Let us  use  the  approach  based  on  the  dimensional regularization
\c{Wilson,Collins}.
It is based on considering integrals \r{p9}  at  arbitrary
dimensionality of  space-time.  Expression  \r{p9}  appears  to  be  a
meromorphic function of $d$.  Substracting the poles corresponding  to
sufficiently small positive integer values of $d$,  we obtain a finite
expression.

Formally, one has
$$
I^{s,0} = \frac{1}{\Gamma(s/2)} \int_0^{\infty} d\alpha \alpha^{s/2-1}
\int d{\bf       k}       e^{-\alpha      ({\bf      k}^2+m^2)}      =
\frac{\pi^{d/2}}{\Gamma(s/2)} \frac{\Gamma(\frac{s-d}{2})}{m^{s-d}}.
$$
If $\frac{s-d}{2} = -N$ is a nonpositive integer  number,  one  should
modify the  definition  of $I^{s,0}$.  Change $d\to d-2{\varepsilon}$.
One finds:
\beb
I^{s,0} =            \frac{\pi^{d/2}}{\Gamma(s/2)             m^{s-d}}
\frac{\Gamma(1+{\varepsilon}) (\pi        m^2)^{-{\varepsilon}       }
}{(-N+{\varepsilon}) ... (-1+{\varepsilon}){\varepsilon} }
\crr
\simeq \frac{\pi^{d/2} (-1)^N}{\Gamma(s/2) m^{s-d} N! {\varepsilon}} (
1+ {\varepsilon} (-ln (\pi m^2) + \Gamma'(1) + 1 + ...  + N^{-1}))  +
O({\varepsilon}).
\eeb
In the $MS$ renormalization scheme \c{Collins}, one should omit the term
$O({\varepsilon}^{-1})$. There    is    also    an     $\overline{MS}$
renormalization scheme  in  which one omits also a fixed term of order
$O(1)$. Let us omit the term $-ln( \pi m^2) + \Gamma'(1)$.  We  obtain
the following renormalized value of the integral:
$$
I^{s,0}_{ren} =          \frac{\pi^{d/2}}{\Gamma(s/2)         m^{s-d}}
\frac{(-1)^N}{N!} (1+ ... + 1/N),
$$
provided that $N= \frac{d-s}{2}$ is a nonnegative integer number.

Therefore, we have defined the renormalized
trace of an asymptotic expansion  of  a
Weyl symbol  by formaula \r{p5}, provided
that the coefficient functions are polynomials in ${\bf k}/\omega_{\bf
k}$.

Let us investigate properties of the renormalized trace.
Some properties are direct corollaries of definition \r{p5}.

{\bf Lemma 3.7.} {\it  The following properties are satisfied:\\
(i) $Tr_R (\underline{A} + \lambda \underline{B}) = 0$;\\
(ii) $Tr_R \frac{\partial \underline{A}}{\partial k_i} = 0$;
$Tr_R \frac{\partial \underline{A}}{\partial x_i} = 0;$\\
(ii) Let $E-\lim_{n\to\infty} \underline{A}_n = \underline{A}$.
Then $\lim_{n\to\infty} Tr_R \underline{A}_n = Tr_R \underline{A}$. \\
(iv) Let $deg \underline{A} > d$. Then $Tr_R \underline{A} = Tr A$.
}

{\bf Corollary.} The property AP9 is satisfied.

Let us check that $Tr_R (\underline{A} * \underline{B} - \underline{B}
* \underline{A}) = 0$. First of all, prove the following statement.

{\bf Lemma 3.8.}
$Tr_R \underline{A} * \underline{B}
= Tr_R \underline{A}\underline{B}$.

{\bf Proof.} Making use of eq.\r{k8}, we find
\beb
(A*B)({\bf x},{\bf k}) - (AB)({\bf x},{\bf k}) =
\crr
\int \frac{d{\bf p}_1
d {\bf  p}_2  d{\bf  y}_1  d{\bf  y}_2}{(2\pi)^{2d}}  \int_0^1 d\alpha
\frac{\partial}{\partial \alpha}
[A({\bf x}+{\bf y}_1; {\bf k}+ \alpha
\frac{{\bf p}_2}{2})   B({\bf  x}+  {\bf  y}_2,  {\bf  k}_2  -  \alpha
\frac{{\bf p}_1}{2})] e^{-i{\bf p}_1{\bf y}_1 - i {\bf p}_2 {\bf y}_2}
=
\crr
- \frac{i}{2}
\int \frac{d{\bf p}_1
d {\bf  p}_2  d{\bf  y}_1  d{\bf  y}_2}{(2\pi)^{2d}}  \int_0^1 d\alpha
\left[
\frac{\partial}{\partial k^i}
A({\bf x}+{\bf y}_1; {\bf k}+ \alpha \frac{{\bf p}_2}{2})
\frac{\partial}{\partial x^i}
B({\bf  x}+  {\bf  y}_2,  {\bf  k}_2  -  \alpha
\frac{{\bf p}_1}{2}) -
\right. \crr \left.
\frac{\partial}{\partial x^i}
A({\bf x}+{\bf y}_1; {\bf k}+ \alpha \frac{{\bf p}_2}{2})
\frac{\partial}{\partial k^i}
B({\bf  x}+  {\bf  y}_2,  {\bf  k}_2  -  \alpha
\frac{{\bf p}_1}{2})
\right]
e^{-i{\bf p}_1{\bf y}_1 - i {\bf p}_2 {\bf y}_2}
= - \frac{i}{2} \frac{\partial C^i({\bf x},{\bf k})}{\partial k^i}
\eeb
with
\beb
C^i({\bf x},{\bf k}) =
\int \frac{d{\bf p}_1
d {\bf  p}_2  d{\bf  y}_1  d{\bf  y}_2}{(2\pi)^{2d}}  \int_0^1 d\alpha
\left[
A({\bf x}+{\bf y}_1; {\bf k}+ \alpha \frac{{\bf p}_2}{2})
\frac{\partial}{\partial x^i}
B({\bf  x}+  {\bf  y}_2,  {\bf  k}_2  -  \alpha
\frac{{\bf p}_1}{2}) -
\right. \crr \left.
B({\bf  x}+  {\bf  y}_2,  {\bf  k}_2  -  \alpha
\frac{\partial}{\partial x^i}
A({\bf x}+{\bf y}_1; {\bf k}+ \alpha \frac{{\bf p}_2}{2})
\frac{{\bf p}_1}{2})
\right]
e^{-i{\bf p}_1{\bf y}_1 - i {\bf p}_2 {\bf y}_2}.
\eeb
One also has
$$
\check{A}*\check{B} -    \check{A}    \check{B}    =   -   \frac{i}{2}
\frac{\partial \check{C}^j}{\partial k^j}
$$
with
\beb
\check{C}^j({\bf x},{\bf  k}) = \sum_{s=0}^{\infty} \sum_{l_1l_2\ge 0;
l_1+l_2 = s}  \frac{(-i)^{l_1}}{2^{l_1}  l_1!}  \frac{i^{l_2}}{2^{l_2}
l_2!} \frac{1}{l_1+l_2+1}
\left[
\frac{\partial^{l_1+l_2} \check{A}}{\partial   k^{i_1}   ...  \partial
k^{i_{l_1}} \partial x^{j_1} ... \partial x^{j_{l_2}}}
\frac{\partial}{\partial x^i}
\frac{\partial^{l_1+l_2} \check{B}}{\partial   k^{j_1}   ...  \partial
k^{j_{l_2}} \partial x^{i_1} ... \partial x^{i_{l_1}} }
\right. \crr \left. -
\frac{\partial^{l_1+l_2} \check{B}}{\partial   k^{j_1}   ...  \partial
k^{j_{l_2}} \partial x^{i_1} ... \partial x^{i_{l_1}}  }
\frac{\partial}{\partial x^i}
\frac{\partial^{l_1+l_2} \check{A}}{\partial   k^{i_1}   ...  \partial
k^{i_{l_1}} \partial x^{j_1} ... \partial x^{j_{l_2}}}.
\right]
\eeb
Analogously to  Appendix C,  one finds that $(C^j,  \check{C}^j)\equiv
\underline{C}^j$ is an asymptotic  expansion  of  a  Weyl  symbol.  It
follows from      lemma     3.7     that     $Tr_R     \frac{\partial
\underline{C}^j}{\partial k^j} = 0$. We obtain statement of lemma 3.8.

{\bf Lemma 3.9.} {\it For $deg \underline{B} \ge 2$,
$Tr_R x^k\omega_{\bf k} * \underline{B} =
Tr_R x^k\omega_{\bf k} \underline{B}$
and
$Tr_R \omega_{\bf k} * \underline{B} =
Tr_R \omega_{\bf k} \underline{B}$.
}

The proof is analogous.

{\bf Corollary 1.} {\it The following relations are satisfied:\\
1. $Tr_R    (\underline{A}   *   \underline{B}   -   \underline{B}   *
\underline{A}) = 0$; \\
2. $Tr_R    (x^k \omega_{\bf k} * \underline{B}   -   \underline{B}   *
x^k \omega_{\bf k})  = 0$; \\
3. $Tr_R    (\omega_{\bf k} * \underline{B}   -   \underline{B}   *
\omega_{\bf k})  = 0$.\\
}

{\bf Corollary 2.} {\it Property \r{trace} is satisfied.}

Thus, we  have constructed functionals
$Tr_R x^k\hat{\Gamma} \equiv Tr_R x^k \underline{\Gamma}$
and
$Tr_R \hat{\Gamma}
\equiv Tr_R \underline{\Gamma}$ such that  properties  \r{trace}
and P6'(b) are satisfied.

Note that the "finite renormalization" \c{BS} can be also be made. One
can add quantities $\Delta Tr_R x^k\hat{\Gamma}$
and $\Delta Tr_R \hat{\Gamma}$ to
renormalized traces in such a way that
\beb
\delta^P_k \Delta Tr_R \hat{\Gamma} = 0; \qquad
\delta^M_{kl} \Delta Tr_R\hat{\Gamma} = 0;
\crr
\delta^P_l \Delta Tr_R x^k \hat{\Gamma}
= - \delta^{kl} \Delta Tr_R \hat{\Gamma};
\qquad
\delta^M_{kl} \Delta Tr_R  x^k\hat{\Gamma}  =  \delta^{kl}
\Delta Tr_R   x^m   \hat{\Gamma}   -
\delta^{mk} \Delta Tr_R x^l \hat{\Gamma};\crr
\delta_l^B \Delta Tr_R x^k\hat{\Gamma}
- \delta_k^B \Delta Tr_R x^l \hat{\Gamma} = 0; \crr
\delta_l^B \Delta Tr_R \hat{\Gamma}
- \delta^H \Delta Tr_R x^l \hat{\Gamma} = 0.
\eeb
This corresponds to the possibility of adding the finite
one-loop counterterm to the Lagrangian.

\section{Semiclassical field}

An important feature of QFT is a notion of field.  In this section  we
introduce the  notion  of a semiclassical field and check its Poincare
invariance.

\subsection{Definition of semiclassical field}

First of  all,  introduce  the  notion  of   a   semiclassical   field
$\tilde{\phi}({\bf x},t:X)$  in  the
functional Schrodinger representation.  At $t=0$, this is the operator
of multiplication by $\phi({\bf x})$. For arbitrary $t$, one has
$$
\tilde{\phi}({\bf x},t:X)  = \tilde{U}_{-t}(X\gets u_tX) \phi({\bf x})
\tilde{U}_t (u_tX  \gets X),
$$
where $\tilde{U}_t(u_tX\gets X)$  is  the  operator  transforming  the
initial condition  for the Cauchy problem for eq.\r{4} to the solution
to the Cauchy problem.

The field  operator  in  the  Fock  representation  is  related   with
$\tilde{\phi}$ by the transformation \r{6*},
$$
\hat{\phi}({\bf x},t:X) = V_X^{-1} \tilde{\phi}({\bf x},t:X) V_X.
$$
Making use of eq.\r{b0}, one finds
\be
\hat{\phi}({\bf x},t:X)  =  (U_H^t(X))^{-1} \hat{\phi}({\bf x},  u_tX)
U_H^t(X)
\l{r0}
\ee
Here $\hat{\phi}({\bf  x}:X)  = i (\Gamma^{-1/2} (A^+-A^-))({\bf x})$,
while
$$
U_H^t(X) \equiv  U_{a,\Lambda}  (u_{a,\Lambda}X   \gets   X),   \qquad
\Lambda=1, {\bf a} =0, a^0=-t.
$$

Let us define $\hat{\phi}$ mathematically as an operator distribution.

Let $S({\bf R}^d)$ be a space of complex  smooth  functions  $u:  {\bf
R}^d \to {\bf C}$ such that
$$
||u||_{l,m} = \max_{\alpha_1 + ... + \alpha_d \le l} \sup_{{\bf x} \in
{\bf R}^d} (1+|{\bf x}|)^m
|\frac{\partial^{\alpha_1 + ...  + \alpha_d}}{\partial x^{\alpha_1} ...
\partial x^{\alpha_d}} u({\bf x}) |
\to_{k\to\infty} 0.
$$
We say   that   the   sequence   $\{u_k\}  \in  S({\bf  R}^n)$,  $k  =
\overline{1,\infty}$ tends to zero if $||u_k||_{l,m}  \to_{k\to\infty}
0$ for all $l,m$.

Denote ${\cal D} = \{ \Psi \in {\cal F} | ||A^+TA^- \Psi|| < \infty \}$
(cf. \c{Shvedov}).

{\bf Definition 4.1.} (cf.\c{A3}).
{\it 1. An operator distribution $\phi$ defined on ${\cal D} \in {\cal
F}$ is  a  linear  mapping taking functions $f\in S({\bf R}^d)$ to the
linear operator $\phi[f]: {\cal D} \to {\cal F}$,
$$
\phi: f\in S({\bf R}^d) \mapsto \phi[f]: {\cal D} \to {\cal F},
$$
such that   $||\phi[f_n]   \Phi||   \to_{n\to\infty}   0$    if    $f_n
\to_{n\to\infty} 0$. \crr
2. A sequence of operator distributions $\phi_n$ is called  convergent
to the operator distribtion $\phi$ if
$$
||\phi_n[f] \Phi - \phi[f] \Phi|| \to_{n\to\infty} 0
$$
for all $\Phi \in {\cal D}$, $f\in S({\bf R}^d)$.
}

We will write
$$
\phi[f] \equiv \int d{\bf x} \phi({\bf x}) f({\bf x}), \qquad {\bf x }
\in {\bf R}^d.
$$

Consider the mapping $f\mapsto \phi_t\{f\}$,  $f\in S({\bf  R}^d)$  of
the form
$$
\phi_t\{f:X\} = \int d{\bf x} \hat{\phi}({\bf x},t:X) f({\bf x}).
$$

It follows from the results of \c{Shvedov} that 
$\phi_t$  is  an  operator  distribution  being
continous with respect to $t$.

Consider the mapping $f\mapsto \phi[f]$,  $f \in  S({\bf  R}^{d+1})$ of
the form
$$
\phi[f:X] = \int dt \phi_t\{f(\cdot,t):X\}.
$$

Analogously, note also that
$\phi$ is an operator distribution.

\subsection{Poincare invariance of the semiclassical field}

\subsubsection{Algebraic properties}

To check the property  of  Poincare  invariance,  notice  that  it  is
sufficient to  check  it  for  partial  cases:  spatial  translations,
rotations, evolution,  boost, since any Poincare transformation can be
presented as a composition of these transformations.  Let $g_B(\tau) =
(a(\tau),\Lambda(\tau))$ be  a  one-parametric  subgroup  of  Poincare
group corresponding  to  the element $B$ of the Poincare algebra.  The
Poincare invariance property can be rewritten as
\be
\hat{\phi} [f:X]    =     (U_B^{\tau}(X))^{+}     \phi[v_{g_B(\tau)}f:
U_{g_B(\tau)} X] U_B^{\tau}[X],
\l{r1}
\ee
where
$$
(v_{g_B(\tau)} f)(x) = f(\Lambda^{-1}(\tau) (x-a(\tau))).
$$
Obviously, $v_{g_1} v_{g_2} = v_{g_1g_2}$.

Let us check relation \r{r1}.  It is convenient to  reduce  the  group
property to an algebraic property.  The formal derivative with respect
to $\tau$ of the right-hand side of eq.\r{r1} is
\be
(U_B^{\tau}(X))^+ \{
i [H(B:u_{g_B(\tau)}X) ; \phi[v_{g_B(\tau)}f: u_{g_B(\tau)}X]]
+ \frac{\partial}{\partial\tau} \phi[v_{g_B(\tau)}f: u_{g_B(\tau)}X]
\} U_B^{\tau}(X)
\l{r2}
\ee
If the quantity \r{r2} vanishes, the property \r{r1} will be satisfied
since it  is  obeyed  at  $\tau=0$.  Making  use of the group property
$g_B(\tau+\delta\tau)= g_B(\delta  \tau)  g_B(\tau)$,  we  find   that
vanishing of expression \r{r2} is equivalent to the property:
\be
\frac{\partial}{\partial \tau} |_{\tau=0}         \phi[v_{g_B(\tau)}f:
u_{g_B(\tau)} X] - i [\phi[f:X] ; H(B:X)] = 0.
\l{r3}
\ee
We obtain the following lemma.

{\bf Lemma 4.1.} {\it Let the bilinear form \r{r3} vanish on $\cal D$.
Then the property \r{r1} is satisfied on $\cal D$.}

{\bf Proof.} Consider the matrix element
$$
\chi^{\tau} =     (U_B^{\tau}(X)     \Psi_1,      \phi[v_{g_B(\tau)}f:
u_{g_B(\tau)}X] U_B^{\tau}(X)   \Psi_2)   -  (\Psi_1,  \hat{\phi}[f:X]
\Psi_2),
$$
where $\Psi_1, \Psi_2 \in {\cal D}$. Show it to be differentiable with
respect to $\tau$. Let us check that 
for  $\Psi  \in  {\cal  D}$, the vector
$\phi[v_{g_B(\tau)}f: u_{g_B(\tau)}X] \Psi$  is  strongly  continously
differentiable with respect to $\tau$. 

One has:
\beb
\frac{\phi[v_{g_B(\tau+ \delta \tau)}f: u_{g_B(\tau+\delta \tau)}X]
- \phi[v_{g_B(\tau)}f: u_{g_B(\tau)}X]}{\delta \tau} \Psi =
\phi[ \frac{v_{g_B(\delta\tau)}-1}{\delta \tau}
v_{g_B(\tau)}f: u_{g_B(\tau+\delta\tau)}X] \Psi + \crr
\frac{\phi[v_{g_B(\tau)}f: u_{g_B(\delta \tau) g_B(\tau)}X]
- \phi[v_{g_B(\tau)}f: u_{g_B(\tau)}X]}{\delta \tau} \Psi.
\eeb
It follows from \c{Shvedov} that
the first term tends to
$\phi[ \frac{\partial}{\partial t}|_{t=0} v_{g_B(t)}
v_{g_B(\tau)}f: u_{g_B(\tau)}X]\Psi$, while
the second    term    tends    to    $\delta[B]   \phi[v_{g_B(\tau)}f:
u_{g_B(\tau)}X]\Psi$.

Notice that
\beb
\frac{\chi^{\tau+\delta \tau} - \chi^{\tau}}{\delta \tau}  =
(\phi[v_{g_B(\tau+ \delta \tau)}f: u_{g_B(\tau + \delta \tau)}X]
U_B^{\tau+\delta \tau}(X) \Psi_1;
\frac{U_B^{\tau+\delta\tau}(X) - U_B^{\tau}}{\delta \tau} \Psi_2)
\crr
+
(U_B^{\tau+\delta\tau}(X) \Psi_1;
(\phi[v_{g_B(\tau+ \delta\tau)}f: u_{g_B(\tau+\delta \tau)}X]
- \phi[v_{g_B(\tau)}f: u_{g_B(\tau)}X]) U_B^{\tau} (X) \Psi_2)
\crr
+
((U_B^{\tau+\delta\tau}(X) - U_B^{\tau}(X)) \Psi_1;
\phi[v_{g_B(\tau)}f: u_{g_B(\tau)}X] U_B^{\tau}(X) \Psi_2).
\eeb
This quantity tends as $\delta \tau\to 0$ to the matrix element of the
bilinear form \r{r2} and vanishes under condition \r{r3}. Lemma 4.1 is
proved.

\subsubsection{Check of invariance}

One should  check  property  \r{r1}  for  spatial   translations   and
rotations, evolution and boost transformations.

For spatial translations and rotations, property \r{r1} reads:
\be
\hat{\phi}({\bf x},t:X)  =  U^{-1}_{0,{\bf a},L} \hat{\phi} (L{\bf x}+
{\bf a}, t: u_{0,{\bf a},L}X) U_{0,{\bf a},L}
\l{r4}
\ee
It follows  from  commutativity  of  $U_{0,{\bf  a},L}$  and $U_t$ and
table 3 that property \r{r4} is satisfied.

For evolution operator, property \r{r1} is rewritten as:
\be
\hat{\phi}({\bf x},t:X)   =    (U_H^{\tau}(X))^{-1}    \hat{\phi}({\bf
x},t-\tau:u_{\tau}X) U_H^{\tau}(X)
\l{r5}
\ee
Relation \r{r5} is a direct corollary of definition \r{r0}  and  group
property for evolution operators.

Consider now the ${\bf n}$-boost transformation. Check property \r{r3}.
It can be presented as
$$
[\breve{B}^k; \hat{\phi}({\bf     x},t;X)]     =     -     i
(x^k
\frac{\partial}{\partial t}   +   t   \frac{\partial}{\partial   x^k})
\hat{\phi}({\bf x},t:X)
$$
or
\bea
[B^k(X) ;   (U_H^t(X))^{-1}   \phi({\bf   x}:u_tX)   U_H^t(X)]   +   i
\delta_B^k\{(U_H^t(X))^{-1}   \phi({\bf   x}:u_tX)   U_H^t(X)\} = \crr
-i (x^k
\frac{\partial}{\partial t}   +   t   \frac{\partial}{\partial   x^k})
(U_H^t(X))^{-1}   \phi({\bf   x}:u_tX)   U_H^t(X)
\l{r6}
\eea
Let us make use of  the property
\be
U_H^t(X) B^k(X)  = i (\delta_B^k U_H^t)(X) + [B^k(u_tX) - t P^k(u_tX)]
U_H^t(X).
\l{r8}
\ee
which can be checked by multiplication by $(U_H^t(X))^{-1}$ and
differentiation with respect to $t$ in a weak sense (cf. \c{Shvedov}).
We take relation \r{r6} to the
form
$$
[\breve{B}^k(Y) -   \breve{P}^k(Y)   t;   \phi({\bf   x}:Y)]   =   x^k
[\breve{H}(Y); \phi({\bf    x}:Y)]    -   it   \frac{\partial\phi({\bf
x}:Y)}{\partial x^k},
$$
where $Y=u_tX$. The property
$$
i \frac{\partial\phi({\bf  x}:Y)}{\partial  x^k}  =   [\breve{P}^k(Y);
\phi({\bf x}:Y)]
$$
is a corollary of relation \r{f7}. The relation
$$
[\breve{B}^k(Y) - x^k \breve{H}(Y); \phi({\bf x}:Y)] = 0
$$
is also checked by direct calculation.

Thus, we have obtained that
the invariance property \r{aa2} is satisfied.

\section{Remarks on composed semiclassical states}

In the soliton quantization theory and in gauge field theories,
the zero-mode problem arises \c{R}. To resolve it, one can 
consider  the
superposition of the "elementary" quantum field  semiclassical  states
\r{2} of the form (cf.\c{MS3})
\be
\psi [\varphi(\cdot)]    =    \int    \frac{d\alpha   }{\lambda^{k/4}}
e^{\frac{i}{\lambda} S(\alpha)}  e^{\frac{i}{\lambda}  \Pi(\alpha;{\bf
x}) [\varphi({\bf   x})   \sqrt{\lambda}  -  \Phi(\alpha,  {\bf  x})]}
g(\alpha, \varphi(\cdot) - \frac{\Phi(\alpha,\cdot)}{\sqrt{\lambda}})
\l{w12}
\ee
where $\alpha\in   {\bf  R}^k$,  $S(\alpha)$,  $\Pi(\alpha;{\bf  x})$,
$\Phi(\alpha;{\bf x})$ are smooth functions.  Calculate (formally) the
functional integral for $(\psi,\psi)$:
\beb
(\psi,\psi) = \int \frac{d\alpha d\gamma}{\lambda^{k/2}} \int D\varphi
e^{-\frac{i}{\lambda} S(\alpha)}  e^{-\frac{i}{\lambda}  \Pi(\alpha;{\bf
x}) [\varphi({\bf   x})   \sqrt{\lambda}  -  \Phi(\alpha,  {\bf  x})]}
g^*(\alpha, \varphi(\cdot) - \frac{\Phi(\alpha,\cdot)}{\sqrt{\lambda}})
\crr
e^{\frac{i}{\lambda} S(\gamma)}  e^{\frac{i}{\lambda}  \Pi(\gamma;{\bf
x}) [\varphi({\bf   x})   \sqrt{\lambda}  -  \Phi(\gamma,  {\bf  x})]}
g(\gamma, \varphi(\cdot) - \frac{\Phi(\gamma,\cdot)}{\sqrt{\lambda}})
\eeb
After substitution  $\gamma  =   \alpha   +   \sqrt{\lambda}   \beta$,
$\varphi(\cdot) =      \frac{\Phi(\alpha,\cdot)}{\sqrt{\lambda}}     +
\phi(\cdot)$ we obtain as $\lambda\to 0$:
\bea
(\psi,\psi) \simeq \int d\alpha d\beta
e^{\frac{i}{\sqrt{\lambda}} \beta_s
(\frac{\partial S}{\partial \alpha_s} - \int d{\bf x} \Pi(\alpha,{\bf x})
\frac{\partial \Phi(\alpha,{\bf x})}{\partial \alpha_s}
)}
e^{\frac{i}{2} \beta_s \frac{\partial}{\partial \alpha_s}
(\frac{\partial S}{\partial \alpha_l} - \int d{\bf x} \Pi(\alpha,{\bf x})
\frac{\partial \Phi(\alpha,{\bf x})}{\partial \alpha_l}) \beta_l
}\crr
\int D\phi g^*(\alpha,\phi(\cdot))
e^{i\beta_l \int d{\bf x}
(\frac{\partial\Pi(\alpha,{\bf x})}{\partial\alpha_l} \phi({\bf x})
- \frac{\partial\Phi(\alpha,  {\bf x})}{\partial \alpha_l} \frac{1}{i}
\frac{\delta}{\delta \phi({\bf x})})}
g(\alpha,\phi(\cdot))
\l{w13}
\eea
The condition
\be
\frac{\partial S}{\partial \alpha_s} =
\int d{\bf x} \Pi(\alpha,{\bf x})
\frac{\partial \Phi(\alpha,{\bf x})}{\partial \alpha_s}
\l{w14}
\ee
should be   satisfied.   Otherwise,   the  integral  \r{w13}  will  be
exponentially small as $\lambda\to 0$,  so that state \r{w12} will  be
trivial. Under condition \r{w14}, one has
\be
(\psi,\psi) \to_{\lambda\to 0} \int d\alpha d\beta
\int D\phi g^*(\alpha,\phi(\cdot))
e^{i\beta_l \int d{\bf x}
(\frac{\partial\Pi(\alpha,{\bf x})}{\partial\alpha_l} \phi({\bf x})
- \frac{\partial\Phi(\alpha,  {\bf x})}{\partial \alpha_l} \frac{1}{i}
\frac{\delta}{\delta \phi({\bf x})})}
g(\alpha,\phi(\cdot))
\l{w15}
\ee

To specify   the   composed  semiclassical  state  in  the  functional
representation, one should:

(i) specify the smooth  functions  $(S(\alpha),  \Pi(\alpha,{\bf  x}),
\Phi(\alpha,{\bf x}))  \equiv X(\alpha)$ obeying eq.\r{w14} (determine
the $k$-dimensional isotropic manifold in  the  extended  phase  space
$\cal X$);

(ii) specify         the         $\alpha$-dependent         functional
$g(\alpha,\phi(\cdot))$.

The inner  product  of  composed  semiclassical  states  is  given  by
expression \r{w15}.

Since the inner product \r{w15} may vanish for nonzero $g$, one should
factorize the space of composed semiclassical states. Such functionals
$g$ that obey the property
\be
\int d\alpha
( g^*(\alpha,\cdot)
\prod_l \delta[\int d{\bf x}
(\frac{\partial\Pi(\alpha,{\bf x})}{\partial\alpha_l} \phi({\bf x})
- \frac{\partial\Phi(\alpha,  {\bf x})}{\partial \alpha_l} \frac{1}{i}
\frac{\delta}{\delta \phi({\bf x})})]
g(\alpha,\cdot) = 0
\l{w16}
\ee
should be set to be equal to zero, $g\sim 0$.

One can  define  the   Poincare   tarnsformation   of   the   composed
semiclassical state  as  follows.  The  transformation of
$(S(\alpha), \Pi(\alpha,\cdot), \Phi(\alpha,\cdot))$ is
$u_{a,\Lambda}(S(\alpha), \Pi(\alpha,\cdot), \Phi(\alpha,\cdot))$. The
transformation of $g(\alpha,\phi(\cdot))$ is
$$
\tilde{U}_{a,\Lambda} (u_{a,\Lambda}(S,\Pi,\Phi)  \gets  (S,\Pi,\Phi))
g(\alpha,\phi(\cdot)).
$$
One should  check  that  the  inner product entering to eq.\r{w16} is
invariant under Poincare transformations.  This will also  imply  that
equivalent states are taken to equivalent.

Since the  functional  Schrodinger representation is not well-defined,
let us consider the Fock representation.  One should then specify  the
$\alpha$-dependent Fock   vector   $Y(\alpha)=  V^{-1}g(\alpha,\cdot)$
instead of the $\alpha$-dependent  functional  $g(\alpha,\phi(\cdot)$.
Making use of formulas \r{6*},  we find that the inner product \r{w15}
takes the form
\bea
\left(
\left(
\matrix{
\Lambda^k \crr Y(\cdot)
}
\right),
\left(
\matrix{
\Lambda^k \crr Y(\cdot)
}
\right)
\right)
= \int d\alpha d\beta
(Y(\alpha), e^{ \beta_s \int d{\bf x} (B_s(\alpha,{\bf x}) A^+({\bf
x}) - B_s^*(\alpha,{\bf x}) A_s^-({\bf x}))} Y(\alpha))
\l{w17}
\eea
where
\be
B_s(\alpha,\cdot) =    \hat{\Gamma}^{-1/2}
(\hat{\cal    R}    \frac{\partial
\Phi(\alpha,\cdot)}{\partial \alpha_s}        -         \frac{\partial
\Pi(\alpha,\cdot)}{\partial \alpha_s}),
\l{w24a}
\ee
$\hat{\Gamma} = \hat{\Gamma}(\Phi(\alpha,\cdot), \Pi(\alpha,\cdot))$,
$\hat{\cal R}= \hat{\cal R}(\Phi(\alpha,\cdot), \Pi(\alpha,\cdot))$.
If the isotropic manifold $(\Phi(\alpha,\cdot), \Pi(\alpha,\cdot))$ is
non-degenerate, the  functions  $B_s(\alpha,{\bf  x})$  are   linearly
independent.

The Poincare transformation of the composed semiclassical state
\beb
\left(
\matrix{
\{ X(\alpha) \} \crr Y(\alpha)
}
\right)
\eeb
is
\beb
\left(
\matrix{
\{ u_{a,\Lambda}X(\alpha) \} \crr
U_{a,\Lambda} (u_{a,\Lambda}X(\alpha) \gets X(\alpha))
Y(\alpha)
}
\right)
\eeb
Consider the quantity
\bea
(B_s,B_l) -  (B_l,B_s)  =  (\frac{\partial\Phi}{\partial   \alpha_s}),
(\hat{\cal R}^*   -   \hat{\cal  R})  \hat{\Gamma}^{-1}
\frac{\partial\Pi}{\partial
\alpha_l}) - ( \frac{\partial\Phi}{\partial \alpha_l},  (\hat{\cal R}^*  -
\hat{\cal R}) \hat{\Gamma}^{-1} \frac{\partial\Pi}{\partial \alpha_s})
= \crr
i \int d{\bf x} \left(
\frac{\partial\Phi(\alpha,{\bf x})}{\partial\alpha_l}
\frac{\partial\Pi(\alpha,{\bf x})}{\partial\alpha_s} -
\frac{\partial\Phi(\alpha,{\bf x})}{\partial\alpha_s}
\frac{\partial\Pi(\alpha,{\bf x})}{\partial\alpha_l}
\right).
\l{w24}
\eea
Differentiating \r{w14}  with  respect  to $\alpha_l$,  we obtain that
quantity \r{w24} vanishes. Thus, operators
$\beta_s \int d{\bf x} (B_s(\alpha,{\bf x}) A^+({\bf
x}) - B_s^*(\alpha,{\bf x}) A_s^-({\bf x}))$ commute each other.

It follows  from  the  resulats  of \c{Shvedov} that the inner product
\r{w24a} is  correctly  defined,  while  Poincare  transformations  of
composed semcialssical  states satisfy the group property and conserve
the inner product \r{w24a}.

\section{Conclusions}

In this  paper  a  notion  of  a  semiclassical  state  is introduced.
"Elementary" semiclassical state are specified by a set $(X,\Psi)$  of
classical  field  configuration $X$ (point on the infinite-dimensional
manifold $\cal X$,  see section 2  and  subsection  3.2)  and  element
$\Psi$  of  the space $\cal F$.  Set of all "elementary" semiclassical
states may be viewed as a semiclassical bundle.

The physical  meaning  of classical field $X$ is evident.  Discuss the
role of $\Psi$. In the soliton quantization language \c{DHN,R} $\Psi$
specifies whether  the  quantum  soliton  is  in the ground or excited
state. For the Gaussian approach \c{G1,G2,G3,G4}, $\Psi$ specifies the
form of the Gaussian functional,  while for QFT in the strong external
classical field \c{GMM,BD} $\Psi$ is a state of a quantum field in the
classical background.

The "composed" semiclassical states have been also introduced (section
5). They can be viewed as superpositions of "elementary" semiclassical
states and  are  specified  by  the  functions  $(X(\tau),\Psi(\tau))$
defined on some domain of ${\bf R}^k$ with values on the semiclassical
bundle.

Not arbitrary  superposition  of  elementary  semiclassical  states is
nontrivial. The  isotropic  condition  \r{w14}  should  be  satisfied.
Moreover, the  inner  product  of  the "composed" semiclassical states
(eq.\r{w17}) is degenerate, so that there is a "gauge freedom" \r{w16}
in specifying composed semiclassical states.

The composed   semiclassical   states  are  used  \c{MS3}  in  soliton
quantization, since there are translation zero modes and solitons  can
be shifted. They are useful if there are conserved integrals of motion
like charges.  The  correspondence  between  composed  and  elementary
semiclassical states in QFT resembles the relationship between WKB and
wave packet approximations in quantum mechanics.

An important feature of QFT is the property of Poincare invariance. In
this paper  an  explicit  check  of  this  property  is  presented for
semiclassical QFT.  The Poincare  transformations  of  elementary  and
composed semiclassical states have been constructed as follows. First,
the simplest Poincare transformations like  spatial  translations  and
rotations,  evolution  and  boost  are  considered.  The infinitesimal
transformations  are  investigated,  the  Lie  algebraic   commutation
relations  have  been  checked  and  the  group  properties  have been
justified.

For the  "composed"  states,  conservation  of  the  degenerate  inner
product and  isotropic  condition  under  Poincare transformation have
been checked.

An important feature of QFT is a notion of field.  In this paper  this
notion is  introduced for semiclassical QFT.  The property of Poincare
invariance of semiclassical field is checked.

This work  was supported by the Russian Foundation for Basic Research,
projects 99-01-01198 and 01-01-06251.

\appendix

\section{Weyl calculus}

The purpose of this appendix is to investigate some properties of Weyl
symbols of operators which are useful in justification  of  properties
P1-P6.

\subsection{Definition of Weyl symbol}

Firs of  all,  remind  the definition of Weyl symbol of operator (see,
for example,  \c{M1,KM}).  Let $A(x,k)$,  $x,k \in  {\bf  R}^d$  be  a
classical observable  depending on coordinates $x = (x_1,...,x_d)$ and
momenta $k = (k_1,...,k_d)$.  To  specify  the  corresponding  quantum
observable $\hat{A}$ (to "quantize" the observable $A$),
one  should
present it as a superposition of
exponents,
$$
A(x,k) = \int d\alpha d\beta \tilde{A}(\alpha,\beta) e^{i\alpha k +  i
\beta x}
$$
and set
$$
\hat{A} =   \int  d\alpha  d\beta  \tilde{A}(\alpha,\beta)  e^{i\alpha
\hat{k} +  i \beta \hat{x}}
$$
Applying the formula for inverse Fourier transformation, we find
\be
(\hat{A} f)     (x)     =     \int     \frac{d\alpha     dp}{(2\pi)^d}
A(x+\frac{\alpha}{2}; p) e^{-i\alpha p} f(x+\alpha).
\l{k5}
\ee
We denote the operator $\hat{A}$ of the form \r{k5} as $\hat{A}= {\cal
W}(A)$. We will also write $A= \overline{\cal W}(\hat{A})$ if $\hat{A}
= {\cal W}(A)$.

{\bf Definition  A.1.}  {\it The  operator  ${\cal  W}(A)$ is called a Weyl
quantization of  the  function  $A$.  The   function   $\overline{\cal
W}(\hat{A})$ is called as a Weyl symbol of the operator $\hat{A}$.}

\subsection{Some calsses of Weyl symbols}

\subsubsection{Classes ${\cal A}_N$ and ${\cal B}_N$}

For investigations  of QFT ultraviolet divergences,  we are interested
in behavior of Weyl symbols of operators at large values  of  momenta.
Let us    introduce   some   important   spaces.   Let   $\omega_k   =
\sqrt{k^2+m^2}$ for some $m$.

{\bf Definition A.2.} {\it 1.  We say that a smooth function  $A(x,k)$
is of the class ${\cal B}_N$ if and only if the functions
\be
\omega_k^{N+s} \frac{\partial^s   A}{\partial   k^{i_1}  ...  \partial
k^{i_s}}
\l{k6+}
\ee
are bounded for all $s$, $i_1,...,i_s$.
\\
2. Let  $A_n  \in  {\cal  B}_N$,  $n=\overline{1,\infty}$,  $A\in {\cal
B}_N$. We say that ${\cal B}_N-\lim_{n\to\infty} A_n = A$ if and  only
if
$$
\lim_{n\to\infty} \max_{k,x}
\omega_k^{N+s} \frac{\partial^s   (A_n - A)}{\partial   k^{i_1}  ...
\partial k^{i_s}} = 0
$$
for all $s$, $i_1,...,i_s$.
\\
3. We  say  that  a  function $A\in {\cal B}_N$ is of the class ${\cal
A}_N$ if and only if
$$
x_{j_1} ...    x_{j_R}    \frac{\partial}{\partial    x_{s_1}}     ...
\frac{\partial}{\partial x_{s_P}} A \in {\cal B}_N
$$
for all $R$, $P$, $j_1,...,j_R$, $s_1,...,s_P$.
\\
4. Let  $A_n \in {\cal A}_N$,  $A \in {\cal A}_N$.  We say that ${\cal
A}_N-lim_{n\to\infty} A_n = A$ if and only if
$$
{\cal B}_N - \lim_{n\to\infty}
x_{j_1} ...    x_{j_R}    \frac{\partial}{\partial    x_{s_1}}     ...
\frac{\partial}{\partial x_{s_P}} (A_n-A) = 0
$$
for all $R$, $P$, $j_1,...,j_R$, $s_1,...,s_P$.
}

Let us  investigate some properties of introduced classes ${\cal A}_N$
and ${\cal B}_N$.

{\bf Lemma A.1.} {\it 1. ${\cal A}_{N+R} \subseteq {\cal A}$,
${\cal B}_{N+R} \subseteq {\cal B}$ for $R\ge 0$. \\
2. Let ${\cal A}_{N+R} - \lim_{n\to\infty} A_n =A$ and $R\ge 0$.
Then ${\cal A}_{N} - \lim_{n\to\infty} A_n =A$.\\
3. Let ${\cal B}_{N+R} - \lim_{n\to\infty} A_n =A$ and $R\ge 0$.
Then ${\cal B}_{N} - \lim_{n\to\infty} A_n =A$.
}

The proof is obvious:  it is sufficient to notice that $\omega_k^{-R}$
is a bounded function.

{\bf Lemma A.2.} {\it
1. Let $A \in {\cal B}_N$.  Then $\frac{\partial}{\partial k_i} A  \in
{\cal B}_{N+1}$.\\
2. Let $A \in {\cal A}_N$. Then $x_iA \in {\cal A}_N$,
$\frac{\partial A}{\partial       x_i}      \in      {\cal      A}_N$,
$\frac{\partial}{\partial k_i} A \in {\cal A}_{N+1}$, $f(x)A \in {\cal
A}_N$ for smooth bounded function $f(x)$. \\
3. Let ${\cal B}_N-\lim_{n\to\infty} A_n = A$. Then
${\cal B}_{N+1}-\lim_{n\to\infty} \frac{\partial}{\partial k_i} A_n =
\frac{\partial}{\partial k_i} A$.\\
4. Let ${\cal A}_N-\lim_{n\to\infty} A_n =A$. Then
${\cal A}_N- \lim_{n\to\infty} x_iA_n = x_i A$,
${\cal A}_N- \lim_{n\to\infty}
\frac{\partial A_n}{\partial       x_i}   =
\frac{\partial A}{\partial       x_i}$,
${\cal A}_{N+1}-\lim_{n\to\infty}
\frac{\partial}{\partial k_i} A_n =
\frac{\partial}{\partial k_i} A$,
${\cal A}_{N}-\lim_{n\to\infty}
f(x)A_n = f(x) A$
for smooth bounded function $f(x)$. \\
}

The proof is also obvious.

{\bf Lemma A.3.} {\it Let $A_1  \in  {\cal  B}_{N_1}$,  $A_2\in  {\cal
B}_{N_2}$. Then $A_1A_2 \in {\cal B}_{N_1+N_2}$. }

{\bf Proof.} It is sufficient to check that the expression
$$
\omega_k^{N_1} \omega_k^{N_2} \omega_k^s \frac{\partial^s}
{\partial k_{i_1} ... \partial k_{i_s} } (A_1A_2)
$$
is bounded. This statement is a corollary of properties $A_1 \in {\cal
B}_{N_1}$, $A_2 \in {\cal B}_{N_2}$ and formula
$$
\frac{\partial}{\partial k_i}  (fg)  =  \frac{\partial}{\partial  k_i}
f\cdot  g + f \cdot \frac{\partial}{\partial k_i} g.
$$
Lemma A.3 is proved.

{\bf Lemma A.4.} {\it The following properties are satisfied:
$k_i \in {\cal B}_{-1}$, $\omega_k^{\alpha} \in {\cal B}_{-\alpha}$.
}

{\bf Proof.} Since $|k_i/\omega_k| <1$,  we obtain
the property    $k_i    \in   {\cal   B}_{-1}$.   For   the   function
$\omega_k^{\alpha}$, one has
\be
\frac{\partial}{\partial k_{i_1}}
...
\frac{\partial}{\partial k_{i_s}}         \omega_k^{\alpha}          =
\omega_k^{\alpha} {\cal P}(k_i/\omega_k)
\l{k7}
\ee
where $\cal P$ is a polynomial in $k_i/\omega_k$.  Property \r{k7}  is
checked by  induction.  Therefore,  functions  \r{k6+} are bounded for
$N=1$. Lemma A.4 is proved.

{\bf Lemma A.5.} {\it
1. Let $A \in {\cal B}_N$. Then
$$
k_iA \in   {\cal  B}_{N-1},  \qquad  \omega_k^{-\alpha}  A  \in  {\cal
B}_{N+\alpha}, \qquad  \frac{\partial  A}{\partial  k_i}   \in   {\cal
B}_{N+1}.
$$
2. Let $A \in {\cal A}_N$. Then
$$
k_iA \in   {\cal  A}_{N-1},  \qquad  \omega_k^{-\alpha}  A  \in  {\cal
A}_{N+\alpha}, \qquad  \frac{\partial  A}{\partial  k_i}   \in   {\cal
A}_{N+1}.
$$
}

{\bf Proof.} Property 1 is a corollary of lemmas A.2 and A.4. Property
1 implies property 2. Lemma is proved.

{\bf Lemma A.6.} {\it
1. Let ${\cal B}_N-\lim_{n\to\infty} A_n = A$. Then
$$
{\cal  B}_{N-1}-\lim_{n\to\infty} k_iA_n = k_i A, \qquad
{\cal B}_{N+\alpha}-\lim_{n\to\infty}
\omega_k^{-\alpha}  A_n = \omega_k^{-\alpha}  A,
\qquad
{\cal B}_{N+1}- \lim_{n\to\infty}
\frac{\partial  A_n}{\partial  k_i}
= \frac{\partial  A}{\partial  k_i}
$$
2. Let ${\cal A}_N-\lim_{n\to\infty} A_n = A$. Then
$$
{\cal  A}_{N-1}-\lim_{n\to\infty} k_iA_n = k_i A, \qquad
{\cal A}_{N+\alpha}-\lim_{n\to\infty}
\omega_k^{-\alpha}  A_n = \omega_k^{-\alpha}  A,
\qquad
{\cal A}_{N+1}- \lim_{n\to\infty}
\frac{\partial  A_n}{\partial  k_i}
= \frac{\partial  A}{\partial  k_i}.
$$
}

The proof is analogous to the proof of lemma A.3.

{\bf Lemma A.7.} {\it
1. Let $A_1 \in {\cal A}_{N_1}$, $A_2 \in {\cal A}_{N_2}$. Then
$A_1A_2 \in {\cal A}_{N_1+N_2}$.\\
2. Let
${\cal A}_{N_1}- \lim_{n\to\infty} A_{1,n} = A_1$,
${\cal A}_{N_2}- \lim_{n\to\infty} A_{2,n} = A_2$.
Then
${\cal A}_{N_1 + N_2}- \lim_{n\to\infty} A_{1,n} A_{2,n} = A_1 A_2$.
}

The proof is analogous to lemma A.3.

\subsubsection{Properties of operators and symbols}

{\bf Lemma A.8.} {\it
1. Let $A\in {\cal A}_0$.  Then the operator ${\cal W}(A)$  \r{k5}  is
bounded.\\
2. Let ${\cal A}_0-\lim_{n\to\infty} A_n =0$.  Then $\lim_{n\to\infty}
||{\cal W}(A_n)|| = 0$.
}

{\bf Proof} (cf.  \c{MS3}).  Let us obtain an estimation for the  norm
$||\hat{A}||$. One has
$$
\hat{A} =  \int  d\beta  e^{i\beta  \hat{x}}  \int  d\alpha e^{i\alpha
(\hat{k}+ \beta/2)} \tilde{A}(\alpha,\beta).
$$
The estimation  $||\int  d\beta  \hat{F}(\beta)||  \le   \int   d\beta
||\hat{F}(\beta)||$ implies
$$
||\hat{A}|| \le   \int  d\beta  ||\int  d\alpha  e^{i\alpha(\hat{k}  +
\beta/2)} \tilde{A}(\alpha,\beta).
$$
However, for operator $F(\hat{k})$ one has  $||F(\hat{k})||  =  \sup_k
||F(k)||$, since  in  the  momentum representation $F(\hat{k})$ is the
operator of multiplication be $F(k)$. Therefore,
\beb
||\int d\alpha  e^{i\alpha  (\hat{k}+\beta/2)} \tilde{A}(\alpha,\beta)
|| =     \max_k     |\int     d\alpha     e^{i\alpha      (k+\beta/2)}
\tilde{A}(\alpha,\beta)| =   \max_k   |\int   d\alpha   e^{i\alpha  k}
\tilde{A}(\alpha,\beta)| = \crr
\max_k |  \int  \frac{dx}{(2\pi)^d}  A(x,k)
e^{-i\beta x}|      =     \frac{1}{(\beta^2+1)^N}     \max_k     |\int
\frac{dx}{(2\pi)^d (x^2+1)^N} e^{-i\beta  x}  (x^2+1)^N  (-\Delta_x  +
1)^N A(x,k)|.
\eeb
Here $N$ is an arbitrary number such that $N> d/2$. Thus,
$$
||\hat{A}|| \le \frac{1}{(2\pi)^d} \int \frac{d\beta dx}{(\beta^2+1)^N
(x^2+1)^N} \max_{kx} |(x^2+1)^N (-\Delta_x+1)^N A(x,k)|.
$$
The first  statement  is  judtified.  Proof of the second statement is
analogous. Lemma A.8 is proved.

{\bf Lemma A.9.}{\it
1. Let   $A\in   {\cal   A}_N$,  $N>d/2$.  Then  ${\cal  W}(A)$  is  a
Hilbert-Schmidt operator.\\
2. Let   ${\cal   A}_N-\lim_{n\to\infty}   A_n  =  0$,  $N>d/2$.  Then
$\lim_{n\to\infty} ||{\cal W}(A_n)||_2 = 0$.
}

{\bf Proof.} Let us use the property \c{M1,KM}
$$
||\hat{A}||_2^2 = \int \frac{dx dk}{(2\pi)^d} |A(x,k)|^2
$$
whcih can be obtained from definition \r{k5}. One has
$$
||\hat{A}||_2^2 \le       \int      \frac{dx}{(2\pi)^d      (x^2+1)^N}
\frac{dk}{\omega_k^{2N}} \max_{xk} |(x^2+1)^{N/2} \omega_k^N A(x,k)|^2.
$$
The first statement is justified.  Proof of the  second  statement  is
analogous. Lemma A.9 is proved.

\subsection{Properties of *-product}

Remind that the Weyl sumbol of the product of operators
$$
A*B = \overline{\cal W} ({\cal W}(A) {\cal W}(B))
$$
can be presented as \c{M1,KM}
\be
(A*B)(x,k) = \int \frac{d\beta_1 d\beta_2 d\xi_1 d\xi_2}{(2\pi)^{2d}}
A(x+\xi_1, k + \frac{\beta_2}{2}) B(x+\xi_2,  k  -  \frac{\beta_1}{2})
e^{-i\beta_1\xi_1 - i\beta_2\xi_2}
\l{k8}
\ee
Formula \r{k8} can be obtained from definition \r{k5}.

Let us investigate some properties of formula \r{k8}.  Let us find  an
expansion of formula \r{k8} as $|k| \to\infty$. Formally, one has
\beb
A(x+\xi_1, k    +    \frac{\beta_2}{2})    =     \sum_{n_2=0}^{\infty}
\frac{1}{2^{n_2} n_2!}  \frac{\partial^{n_2} A(x+\xi_1,  k) }{\partial
k^{i_1} ... \partial k^{i_{n_2} }} \beta_2^{i_1} ... \beta_2^{i_{n_2}};
\crr
B(x+\xi_2, k    -  \frac{\beta_1}{2})    =     \sum_{n_1 =0}^{\infty}
\frac{(-1)^{n_1}}
{2^{n_1} n_1!}  \frac{\partial^{n_1} B(x+\xi_2,  k) }{\partial
k^{j_1} ... \partial k^{j_{n_1} }} \beta_1^{j_1} ... \beta_1^{j_{n_1}}.
\eeb
Therefore,
\bea
(A*B)(x,k) =   \sum_{n_1n_2=0}^{\infty}  \frac{(-1)^{n_1}}{2^{n_1+n_2}
n_1! n_2!} \int
\frac{d\beta_1 d\beta_2 d\xi_1 d\xi_2}{(2\pi)^{2d}}
e^{-i\beta_1\xi_1 - i\beta_2\xi_2}
\frac{\partial^{n_2} A(x+\xi_1,  k) }{\partial
k^{i_1} ... \partial k^{i_{n_2} }} \beta_2^{i_1} ... \beta_2^{i_{n_2}}
\crr \times
 \frac{\partial^{n_1} B(x+\xi_2,  k) }{\partial
k^{j_1} ... \partial k^{j_{n_1} }} \beta_1^{j_1} ... \beta_1^{j_{n_1}}
=
\sum_{n_1n_2 = 0}^{\infty} \frac{i^{n_1-n_2}}{2^{n_1+n_2} n_1! n_2!}
\frac{\partial^{n_1 + n_2} A(x,  k) }{\partial
k^{i_1} ... \partial k^{i_{n_2} }
\partial x^{j_1} ... \partial x^{j_{n_1}} }
\frac{\partial^{n_1 + n_2} B(x,  k) }{\partial
x^{i_1} ... \partial x^{i_{n_2} }
\partial k^{j_1} ... \partial k^{j_{n_1}} }
\l{k8a}
\eea
Denote
$$
(A \stackrel{K}{*} B)(x,k) =
\sum_{n_1n_2 \ge 0, n_1+n_2 \le K}
\frac{i^{n_1-n_2}}{2^{n_1+n_2} n_1! n_2!}
\frac{\partial^{n_1 + n_2} A(x,  k) }{\partial
k^{i_1} ... \partial k^{i_{n_2} }
\partial x^{j_1} ... \partial x^{j_{n_1}} }
\frac{\partial^{n_1 + n_2} B(x,  k) }{\partial
x^{i_1} ... \partial x^{i_{n_2} }
\partial k^{j_1} ... \partial k^{j_{n_1}} }
$$
This is an asymptotic expansion in $1/|k|$ as $|k|\to \infty$.  Let us
estimate an accuracy of the asymptotic series.

Making use of the relation
$$
A(x+\xi_1, k   +   \frac{\beta_2}{2})   -   A(x+\xi_1,k)   =  \int_0^1
d(\alpha_2-1) \frac{\partial}{\partial \alpha_2} A(x+\xi_1, k+\alpha_2
\frac{\beta_2}{2})
$$
and integrating by parts $N_2$ times, we find
\beb
A(x+\xi_1, k     +    \frac{\beta_2}{2})    =    \sum_{n_2=0}^{N_2}
\frac{1}{2^{n_2} n_2!} \frac{\partial^{n_2} A(x+\xi_1,k)  }{  \partial
k^{i_1} ... \partial k^{i_{n_2}} } \beta_2^{i_1} ... \beta_2^{i_{n_2}}
\crr + \int_0^1 d\alpha_2 \frac{(1-\alpha_2)^{N_2}}{2^{N_2+1} N_2!}
\frac{\partial^{N_2+1} A(x+\xi_1, k + \alpha_2 \frac{\beta_2}{2})}
{\partial k^{i_1}  ...  \partial  k^{i_{N_2+1}}  }  \beta_2^{i_1}  ...
\beta_2^{i_{N_2+1}}.
\eeb
Analogously,
\beb
B(x+\xi_2, k     -    \frac{\beta_1}{2})    =    \sum_{n_1=0}^{N_1}
\frac{(-1)^{n_1}}{2^{n_1} n_1!}
\frac{\partial^{n_1} A(x+\xi_2,k)  }{  \partial
k^{i_1} ... \partial k^{i_{n_1}} } \beta_1^{i_1} ... \beta_1^{i_{n_2}}
\crr + \int_0^1 d\alpha_1 \frac{(1-\alpha_1)^{N_1} (-1)^{N_1}}
{2^{N_1+1} N_1!}
\frac{\partial^{N_1+1} B(x+\xi_2, k - \alpha_1 \frac{\beta_1}{2})}
{\partial k^{i_1}  ...  \partial  k^{i_{N_1+1}}  }  \beta_1^{i_1}  ...
\beta_1^{i_{N_1+1}}.
\eeb
Therefore,
\beb
(A*B)(x,k) =
\sum_{n_1= 0}^{N_1}
\sum_{n_2= 0}^{N_2}
\frac{i^{n_1-n_2}}{2^{n_1+n_2} n_1! n_2!}
\frac{\partial^{n_1 + n_2} A(x,  k) }{\partial
k^{i_1} ... \partial k^{i_{n_2} }
\partial x^{j_1} ... \partial x^{j_{n_1}} }
\frac{\partial^{n_1 + n_2} B(x,  k) }{\partial
x^{i_1} ... \partial x^{i_{n_2} }
\partial k^{j_1} ... \partial k^{j_{n_1}} }
\crr + \sum_{n_1   =   0}^{N_1}  r^{(1)}_{n_1N_2}  +  \sum_{n_2=0}^{\infty}
r^{(2)}_{N_1n_2} + R_{N_1N_2}
\eeb
with the following remaining terms,
\beb
r^{(1)}_{n_1N_2} =     \int     \frac{d\beta_1     d\beta_2     d\xi_1
d\xi_2}{(2\pi)^{2d}} e^{-i\beta_1 \xi_1 - i \beta_2 \xi_2}
\int_0^1 d\alpha_2 \frac{(1-\alpha_2)^{N_2}}{2^{N_2+1} N_2!}
\frac{\partial^{N_2+1} A(x+\xi_1, k + \alpha_2 \frac{\beta_2}{2})}
{\partial k^{i_1}  ...  \partial  k^{i_{N_2+1}}  }  \beta_2^{i_1}  ...
\beta_2^{i_{N_2+1}} \crr \times  \frac{(-1)^{n_1} }{2^{n_1} n_1! }
\frac{\partial^{n_1} B(x+\xi_1,   k)}{\partial  k^{j_1}  ...  \partial
k^{j_{n_1}} } \beta_1^{j_1} ... \beta_1^{j_{n_1}};
\eeb
\beb
r^{(2)}_{N_1n_2} =     \int     \frac{d\beta_1     d\beta_2     d\xi_1
d\xi_2}{(2\pi)^{2d}} e^{-i\beta_1 \xi_1 - i \beta_2 \xi_2}
\int_0^1 d\alpha_1 \frac{(1-\alpha_1)^{N_1} (-1)^{N_1+1}}
{2^{N_1+1} N_1!}
\frac{\partial^{N_1+1} B(x+\xi_2, k - \alpha_1 \frac{\beta_1}{2})}
{\partial k^{j_1}  ...  \partial  k^{j_{N_1+1}}  }  \beta_1^{j_1}  ...
\beta_1^{j_{N_1+1}} \crr \times  \frac{1}{2^{n_2} n_2! }
\frac{\partial^{n_2} B(x+\xi_2,   k)}{\partial  k^{i_1}  ...  \partial
k^{i_{n_2}} } \beta_2^{i_1} ... \beta_2^{i_{n_2}};
\eeb
\beb
R_{N_1N_2} =     \int     \frac{d\beta_1     d\beta_2     d\xi_1
d\xi_2}{(2\pi)^{2d}} e^{-i\beta_1 \xi_1 - i \beta_2 \xi_2}
\int_0^1 d\alpha_2 \frac{(1-\alpha_2)^{N_2}}{2^{N_2+1} N_2!}
\frac{\partial^{N_2+1} A(x+\xi_1, k + \alpha_2 \frac{\beta_2}{2})}
{\partial k^{i_1}  ...  \partial  k^{i_{N_2+1}}  }  \beta_2^{i_1}  ...
\beta_2^{i_{N_2+1}}
\crr \times \int_0^1 d\alpha_1 \frac{(1-\alpha_2)^{N_2} (-1)^{N_1+1}}
{2^{N_1+1} N_1!}
\frac{\partial^{N_1+1} B(x+\xi_2, k - \alpha_1 \frac{\beta_1}{2})}
{\partial k^{j_1}  ...  \partial  k^{j_{N_1+1}}  }  \beta_1^{j_1}  ...
\beta_1^{j_{N_1+1}}.
\eeb

Let us investigate the remaining terms.

\subsubsection{The $k$-independent case}

{\bf Definition A.3.} {\it We say that the function $f(x)$, $x\in {\bf
R}^d$ is  of  the class $\cal C$ if $f$ is a smooth function such that
for each set $(i_1,...,i_l)$ there exists $m>0$ such that the function
$$
(x^2+1)^{-m} \frac{\partial^l}{\partial x^{i_1} ... \partial x^{i_l}}f
$$
is bounded.
}

Let $A=f(x)$,$f  \in  {\cal  C}$.  Then  the  only  nontrivial term is
$r^{(2)}_{N_10}$ which is taken by integrating by parts to the form
\be
r^{(2)}_{N_10} =      \int      \frac{d\beta_1       d\xi_1}{(2\pi)^d}
e^{-i\beta_1\xi_1} \frac{\partial^{N_1+1}}{\partial     x^{j_1}    ...
\partial x^{j_{N_1+1}} } f(x+\xi_1)
\int_0^1 d\alpha_1 (\frac{i}{2})^{N_1+1} \frac{(1-\alpha_1)^{N_1}}{N_1!}
\frac{\partial^{N_1+1} B(x, k - \alpha_1 \frac{\beta_1}{2})}
{\partial k^{j_1} ... \partial k^{j_{N_1+1}} }.
\l{k9}
\ee
Let us prove some auxiliary statements.

{\bf Lemma A.10.} {\it For some constent $A_1$ the estimation
\be
\omega_k \le A_1 \omega_p \omega_{k-p}
\l{k10}
\ee
is satisfied.
}

{\bf Proof.} Let $p=(\frac{1}{2} + \alpha) k + p_{\perp} $, $\alpha\in
{\bf R}$, $p_{\perp} \perp k$. Then
$$
\frac{\omega_k}{\omega_p \omega_{k-p} } \le \frac{\omega_k}
{\omega_{(1/2+\alpha)k} \omega_{(1/2-\alpha)k}} \equiv f(\alpha,k),
$$
so that it is sufficient to check estimation \r{k10} for $p=\alpha  k$
only. For the function $1/f^2$, one has
$$
\frac{1}{f^2(\alpha,k)} =  \frac{1}{k^2+m^2}
[(\frac{1}{2} + \alpha)^2 k^2 + m^2]
[(\frac{1}{2} - \alpha)^2 k^2 + m^2].
$$
It has the following minimal value
\bea
min_{\alpha}
\frac{1}{f^2(\alpha,k)} =
\left\{
\matrix{
\frac{(k^2/4+m^2)^2}{k^2+m^2}, \qquad k^2<4m^2, \qquad \alpha=0. \crr
\frac{k^2m^2}{k^2+m^2}, \qquad      k^2      >      4m^2,       \qquad
\alpha=\sqrt{\frac{1}{4}- \frac{m^2}{k^2}}.
}
\right.
\l{k10*}
\eea
The quantity \r{k10*} is bounded below. Thus, lemma is proved.

{\bf Corollary.} {\it For $0<\gamma <1$,
$$
\frac{\omega_k}{\omega_p \omega_{k-\gamma p}} \le A_1.
$$
}

{\bf Lemma A.11.} {\it Let $C\in {\cal A}_N$,  $\chi  \in  {\cal  C}$,
$\varphi \in C[0,1]$. Then for
\be
F(x,k) =    \int_0^1   d\alpha   \varphi(\alpha)   \int   \frac{d\beta
d\xi}{(2\pi)^d} e^{-i\beta  \xi}  \chi(x+\xi)  C(x,  k-   \frac{\alpha
\beta}{2})
\l{k11}
\ee
the function $\omega_k^{N}F$ is bounded.
}

{\bf Proof.}  Inserting the identity
\be
e^{-i\beta \xi}  =  (\xi^2  +1)^{-L_1}  (-  \frac{\partial^2}{\partial
\beta^2} + 1)^{L_1} e^{-i\beta \xi}
\l{k11*}
\ee
and integrating by parts, we obtain that
$$
F(x,k) =    \int_0^1   d\alpha   \varphi(\alpha)   \int   \frac{d\beta
d\xi}{(2\pi)^d} \frac{1}{(\xi^2 + 1)^{L_1}} e^{-i\beta\xi} \chi(x+\xi)
\left( 1    -   \frac{\alpha^2}{4}   \frac{\partial^2}{\partial   k^2}
\right)^{L_1} C(x; k - \frac{\alpha \beta}{2}).
$$
For the function $\omega_k^NF$, one has
\bea
\omega_k^N F =  \int_0^1  d\alpha  \varphi(\alpha)  \int  \frac{d\beta
d\xi}{(2\pi)^d} \frac{1}{(\xi^2+1)^{L_1}        \omega_{beta/2}^{L_2}}
\chi(x+\xi) \left( - \frac{1}{4} \frac{\partial^2}{\partial  \xi^2}  +
m^2 \right)^{\frac{L_2+N}{2}}                           e^{-i\beta\xi}
\frac{\omega_k^N}{\omega_{\beta/2}^N} \crr \times
\left( 1    -   \frac{\alpha^2}{4}   \frac{\partial^2}{\partial   k^2}
\right)^{L_1} C(x; k - \frac{\alpha \beta}{2}).
\l{k12}
\eea
Choose $L_2$  to  be  such a number that $\frac{L_2+N}{2}$ is integer,
$L_2>d$. The property $\chi \in {\cal C}$ implies  that  there  exists
such $K$ that
$$
\frac{\partial^m}{\partial \xi_{i_1}    ...   \partial   \xi_{i_n}   }
\chi(x+\xi) = ((x+ \xi)^2 + 1)^K f_{m. i_1...i_m} (x+\xi), \qquad
m = \overline{0, \frac{L_2+N}{2}},
$$
where $f_{m,  i_1...  i_m}$ are bounded functions.  Choose $L_1$ to be
integer and $L_1 > \frac{K+d}{2}$.  Integrating expression \r{k12}  by
parts, making  use of corollary of lemma A.10 and property $C\in {\cal
A}_N$, we obtain that $\omega_k^NF$ is a bounded function.  Lemma A.11
is proved.

{\bf Lemma  A.12.}  {\it  Under  conditions of lemma A.11 $F \in {\cal
A}_N$.}

{\bf Proof.} It is sufficient to consider the functions
\be
\omega_k^{N+I} \frac{\partial^I}{\partial   k_{i_1}    ...    \partial
k_{i_I}} x_{j_1}  ...  x_{j_R}  \frac{\partial}{\partial  x_{s_1}} ...
\frac{\partial}{\partial x_{s_P}}F
\l{k12*}
\ee
which are  expressed  via linear combinations of integrals of the type
\r{k11}. Lemma A.12 is a corollary of lemma A.11.

{\bf Lemma A.13.} {\it
Let ${\cal  A}_N-\lim_{n\to\infty}  C_n  =C$,  $\chi  \in  {\cal  C}$,
$\varphi \in C[0.1]$. Then ${\cal  A}_N-\lim_{n\to\infty}  F_n  =F$.
}

The proof is analogous to lemmas A.11 and A.12.

We obtain therefore the following theorem.

{\bf Theorem A.14.}
{\it
1. Let $f \in {\cal C}$, $B \in {\cal A}_N$. Then
$$
f * B = f \stackrel{K}{*} B + {R}_K
$$
with $R_K \in {\cal A}_{N+K+1}$.\\
2. Let $f\in {\cal C}$, ${\cal A}_N - \lim_{n\to\infty} B_n = 0$. Then
${\cal A}_{N+K+1}
- \lim_{n\to\infty} (f * B_n - f \stackrel{K}{*}B_n) = 0$.
}

\subsubsection{The $x$-independent case}

Let $A=A(k)$,  $A\in {\cal B}_{M_1}$, $B \in {\cal A}_{M_2}$. The only
nontrivial term is taken to the form:
$$
r^{(1)}_{0N_2}(x,k) =                \int_0^1                d\alpha_2
\left(-\frac{i}{2}\right)^{N_2+1} \frac{(1-\alpha_2)^{N_2}     }{N_2!}
\int \frac{d\beta_2 d\xi_2}{(2\pi)^d} e^{-i\beta_2\xi_2}
\frac{\partial^{N_2+1} A(k  +  \frac{\alpha_2  \beta_2}{2}) }{\partial
k^{i_1} ... \partial k^{i_{N_2+1}} } \frac{\partial^{N_2+1} B(x+\xi_2;
k)}{\partial x^{i_1} ... \partial x^{i_{N_2+1}} }.
$$

{\bf Lemma A.15.} {\it $C=C(k)$,  $C \in {\cal  B}_{K_1}$,  $K_1  >0$,
$D\in {\cal A}_{K_2}$, $\varphi \in C[0,1]$. Then for
$$
F(x,k) =    \int_0^1   d\alpha   \varphi(\alpha)   \int   \frac{d\beta
d\xi}{(2\pi)^d} e^{-i\beta \xi} C(k+ \frac{\alpha \beta}{2})  D(x+\xi,
k) \xi_{j_1} ... \xi_{j_m}
$$
the function $\omega_k^{K_1+K_2}F$ is bounded.
}

{\bf Proof.} Inserting the identity \r{k11*} and integrating by parts,
we obtain that
\beb
F(x,k) =    \int_0^1   d\alpha   \varphi(\alpha)   \int   \frac{d\beta
d\xi}{(2\pi)^d} \frac{1}{(\xi^2+1)^{L_1}}  e^{-i\beta\xi}   D(x+\xi,k)
\left( 1    -   \frac{\alpha^2}{4}   \frac{\partial^2}{\partial   k^2}
\right)^{L_1}
\crr \times
(-\frac{i\alpha}{2})^m \frac{\partial}{\partial k_{j_1}}
... \frac{\partial}{\partial k_{j_m}} C(k + \frac{\alpha \beta}{2}).
\eeb
For the function $\omega_k^{K_1+K_2}F$, one has
\beb
\omega_k^{K_1+K_2} F(x,k)  =  \int_0^1  d\alpha  \varphi(\alpha)  \int
\frac{d\beta d\xi}{(2\pi)^d}                  \frac{1}{(\xi^2+1)^{L_1}
\omega_{\beta/2}^{L_2} } \omega_k^{K_2} D(x+\xi,k)  \left(
- \frac{1}{4} \frac{\partial^2}{\partial \xi^2} + m^2
\right)^{\frac{L_2+K_1}{2}}
\crr
e^{-i\beta\xi}
\frac{\omega_k^{K_1}}{\omega_{\beta/2}^{K_1} }
\left( 1- \frac{\alpha^2}{4} \frac{\partial^2}{\partial k^2}
\right)^{L_1} (-i\frac{\alpha}{2})^m        \frac{\partial^m}{\partial
k_{j_1} ... \partial k_{j_m}} C(k + \frac{\alpha \beta}{2})
\eeb
Integrating by parts for sufficiently large $L_1$,  $L_2$,  making use
of lemmas A.10, we check proposition of lemma A.15.

{\bf Lemma A.16.} {\it Under conditions of lemma  A.15  $F  \in  {\cal
A}_{K_1+K_2}$. }

{\bf Lemma A.17.} {\it Let ${\cal A}_{K_2}-\lim_{n\to\infty} D_n = D$,
$C=C(k)$, $C\in {\cal B}_{K_1}$, $K_1>0$, $\varphi \in C[0,1]$.
Then
${\cal A}_{K_1+K_2}-\lim_{n\to\infty} F_n = F$.
}

The proof  is  analogous  to lemmas A.12 and A.13.  We obtain then the
following theorem.

{\bf Theorem A.18.} {\it
1. Let $A=A(k)$, $A \in {\cal B}_{M_1}$, $B \in {\cal A}_{M_2}$. Then
$$
A*B = A \stackrel{K}{*} B + R_K
$$
with $R_K \in {\cal A}_{M_1+M_2+K+1}$, provided that $K+M_1+1>0$.\\
2. Let $A=A(k)$, $A\in {\cal B}_{M_1}$,
${\cal A}_{M_2}-\lim_{n\to\infty} B_n = B$. Then
$$
{\cal A}_{M_1+M_2+K+1}-\lim_{n\to\infty}
(A*B_n - A \stackrel{K}{*} B_n) =0,
$$
provided that $K+M_1+1>0$.
}

{\bf Remark.}  If  the  proposition  of  theorem A.18 is satisfied for
$K=K_0$, it is satisfied for all $K\le K_0$.  Therefore, the condition
$K+M_1+1>0$ can be omitted.

The following lemma is a corollary of theorem A.18.

{\bf Lemma A.19.}{\it
1. Let $A\in {\cal A}_N$,  $N>d$.  Then ${\cal W}(A)$ is of the  trace
class.\\
2. Let ${\cal A}_N-\lim_{n\to\infty} A_n=0$,  $N>d$.  Then
$\lim_{n\to\infty} Tr  {\cal W}(A_n) = 0$.
}

{\bf Proof.} Consider the operator
$$
\hat{B} = {\cal W}(B) = \hat{\omega}^{N/2} (x^2+1)^{N/2} {\cal W}(A)
$$
with
$$
B = \omega_k^{N/2} * (x^2+1)^{N/2} * A
$$
Since $B \in {\cal  A}_{N/2}$,  ${\cal  W}(B)$  is  a  Hilbert-Schmidt
operator according to lemma A.9. Therefore, ${\cal W}(A)$ is a product
of two Hilbert-Schmidt operators $(x^2+1)^{-N/2}  \hat{\omega}^{-N/2}$
and ${\cal W}(B)$. Thus, ${\cal W}(A)$ is of the trace class.

One also has:
$$
|Tr {\cal  W}(A_n)|  =  |Tr  (x^2+1)^{-N/2}  \hat{\omega}^{-N/2} {\cal
W}(B_n)| \le    ||(x^2+1)^{-N/2}    \hat{\omega}^{-N/2}||_2    ||{\cal
W}(B_n)||_2.
$$
Making use of lemma A.9, we prove lemma A.19.

\subsubsection{The ${\cal A}_N$-case}

Let $A \in {\cal A}_{M_1}$,  $B \in {\cal A}_{M_2}$. The $r$-terms can
be investigated as follows.

1. We substitute $\beta_{1,2}^j e^{-i\beta_{1,2}\xi_{1.2}} \equiv
i \frac{\partial}{\partial \xi_{1,2}^j} e^{-i\beta_{1,2}\xi_{1.2}}$
and integrate the expressions for $r^{(1)}$,  $r^{(2)}$,  $R$ by parts
with respect to $\xi_1$, $\xi_2$.

2. We consider the quantities like
$$
\omega_k^{N_1+N_2+M_1+ M_2+1  + L} \frac{\partial^L}{ \partial k^{i_1}
... \partial k^{i_L}}  x_{j_1}  ...  x_{j_J}
\frac{\partial}{\partial x_{s_1}} ...
\frac{\partial}{\partial x_{s_P}} r
$$
for $r =r^{(1)},  r^{(2)},  R$ and show them to be bounded. We use the
following statement.

{\bf Lemma  A.20.}  {\it  Let  $F  \in  {\cal A}_{K_1}$,  $G \in {\cal
A}_{K_2}$, $K_1,K_2>0$. Then the function
$$
\int \frac{d\beta_1     d\beta_2      d\xi_1      d\xi_2}{(2\pi)^{2d}}
e^{-i\beta_1\xi_1 -  i\beta_2 \xi_2} \omega_k^{K_1+K_2} F(x+\xi_1,  k+
\alpha_2 \frac{\beta_2}{2}) G(x+\xi_2, k - \alpha_1 \frac{\beta_1}{2})
\xi_1^{j_1} ... \xi_1^{j_m}
$$
is uniformly bounded with respect to $\alpha_1, \alpha_2 \in [0,1]$.
}

This lemma is proved analogously to lemmas A.11 and A.15.

3. Analogously to previous subsubsections,
we prove the following theorem.

{\bf Theorem A.21.}
{\it
1. Let $A \in {\cal A}_{M_1}$, $B \in {\cal A}_{M_2}$. Then
$$
A*B = A \stackrel{K}{*} B + R_K
$$
with $R_K \in {\cal A}_{M_1+M_2+K+1}$.\\
2. Let $A_n \in {\cal A}_{M_1}$, $B_n \in {\cal A}_{M_2}$. Then
$$
{\cal A}_{M_1+M_2+K+1}-\lim_{n\to\infty}
(A_n*B_n      -      A_n \stackrel{K}{*} B_n)
=
A*B - A \stackrel{K}{*} B.
$$
}

\subsection{Properties of the exponent}

Let us investigate now the properties of the exponent of the  operator
$\exp {\cal  W}(A)  \equiv  {\cal  W}(*\exp  A)$.  It is convenient to
consider the Fourier transformations of Weyl symbols,
$$
\tilde{A}(\gamma,k) = \int \frac{dx}{(2\pi)^d} e^{-i\gamma x} A(x,k).
$$
Introduce the following norms for Weyl symbols,
\be
||A||_{I,K} =  \max_{J+M+N  \le  K}  \max_{\gamma,  K} |\omega_k^{I+J}
\frac{\partial^J}{\partial k_{j_1} ...  \partial k_{j_J}} \gamma_{m_1}
... \gamma_{m_M} \frac{\partial^N \tilde{A}}{\partial \gamma_{n_1} ...
\partial \gamma_{n_N} }|.
\l{k19}
\ee

{\bf Lemma A.22.} {\it
$A\in {\cal  A}_I$  if  and only if $||A||_{I,K} < \infty$ for all $k=
\overline{0,\infty}$.
}

The proof is obvious.

Let $C=A*B$.  Then  the  Fourier  transformation  $\tilde{C}$  can  be
expressed via $\tilde{A}$ and $\tilde{B}$ as follows,
\be
\tilde{C}(\gamma,k) =   \int    d\alpha    \tilde{A}(\alpha,    k    +
\frac{\gamma-\alpha}{2}) \tilde{B}(\gamma-\alpha,          k         -
\frac{\alpha}{2}).
\l{k20}
\ee
The following estimation is satisfied.

{\bf Lemma  A.23.} {\it
For  arbitrary integer numbers $K$,  $L>d/2$ there
exists such a constant $b_K$ that
\be
||A*B||_{0,K} \le b_K ||A||_{0, K+2L} ||B||_{0,K}.
\l{k21}
\ee
}

To prove estimation \r{k21},  one should use  definition  \r{k19}  and
formula \r{k20}:\\
(i) the derivatives $\partial/\partial \gamma_n$ are applied as
\beb
\frac{\partial}{\partial \gamma_n}     (\tilde{A}(\alpha,     k      +
\frac{\gamma-\alpha}{2}) \tilde{B}(\gamma-\alpha,          k         -
\frac{\alpha}{2})) =  \frac{1}{2}  \frac{\partial  \tilde{A}}{\partial
k_n}(\alpha, k + \crr
\frac{\gamma-\alpha}{2}) \tilde{B}(\gamma-\alpha,  k-
\frac{\alpha}{2}) + \tilde{A}(\alpha, k + \frac{\gamma-\alpha}{2})
\frac{\partial}{\partial \gamma_n}    \tilde{B}(\gamma-\alpha,   k   -
\frac{\alpha}{2});
\eeb
(ii) the   derivatives   $\partial/\partial    k_j$    are    applied
analogously;\\
(iii) the  multiplicators  $\gamma_m$  are  written  as   $\alpha_m   +
(\gamma_m - \alpha_m)$; \\
(iv) the estimations
$$
\omega_k \le C \omega_{\alpha/2} \omega_{k-\alpha/2},
\qquad
\omega_k \le C \omega_{\frac{\gamma - \alpha}{2}}
\omega_{k+ \frac{\gamma - \alpha}{2}}
$$
(lemma A.10) are taken into account. \\
(v) the integrating measure is written as
$$
d\alpha = \frac{d\alpha}{(\alpha^2+1)^L} (\alpha^2+1)^L.
$$
We obtain the estimation \r{k21}.

Consider the Weyl symbol of the exponent
\be
*\exp At -1 = \sum_{n=1}^{\infty} \frac{A^{*n} t^n}{n!}
\l{k22}
\ee
with $A^{*n} = A* ... *A$.

{\bf Lemma A.24.} {\it  Let  $A  \in  {\cal  A}_M$,$M>0$.  Then  the
estimation \r{k22}  is  convergent in the $||\cdot||_{0,K}$-norm.  The
estimation $||*\exp At -1||_{0,K} \le  C_K$  is  satisfied  for  $t\in
[0,T]$.}

{\bf Proof.}  One has
$$
||A^{*n}||_{0,K} \le  b_K^{n-1}  ||A||_{0,K+2L}^{n-1}  ||A||_{0,K} \le
b_K^{n-1} ||A||^n_{0,K+2L}.
$$
Therefore,
$$
||*\exp At -1||_{0,K} \le \sum_{n=1}^{\infty}  \frac{1}{b_K}  \frac{(t
||A||_{0,K+2L} b_K)^n}{n!} \le
\frac{e^{t||A||_{0, K+2L} b_K} - 1}{b_K} \le C_K
$$
on $t \in [0.T]$. Lemma A.24 is proved.

{\bf Lemma A.25.} {\it  Let $A \in {\cal A}_M$, $M>0$. Then
$$
\sum_{m=N}^{\infty} \frac{A^{*m}}{m!} \in {\cal A}_{MN}.
$$
}

{\bf Proof.} One has
\be
\sum_{m=N}^{\infty} \frac{A^{*m}}{m!} = A^{*N} \left(
\frac{1}{N!} + \int_0^1 d\tau
\frac{(1-\tau)^{N-1}}{(N-1)!} (*\exp A\tau -1)
\right)
\l{k23}
\ee
Lemma A.24 implies that
$$
\int_0^1 d\tau  \frac{(1-\tau)^{N-1}}{(N-1)!}  (*\exp  A\tau  -1)  \in
{\cal A}_0.
$$
It follows  from theorem A.21 that the symbol \r{k23} is of the ${\cal
A}_{NM}$-class. Lemma A.25 is proved.

{\bf Lemma A.26.} {\it Let $A_n \in  {\cal  A}_M$,  $M>0$  and
${\cal A}_M-\lim_{n\to\infty} A_n = A$. Then
$$
{\cal A}_{MN}-\lim_{n\to\infty}
\sum_{m=N}^{\infty} \frac{A_n^{*m}}{m!}
= \sum_{m=N}^{\infty} \frac{A^{*m}}{m!}.
$$
}

{\bf Proof.}   Because   relation  \r{k23}  and  theorem  A.21  it  is
sufficient to prove that
\be
{\cal A}_0-\lim_{n\to\infty}  \int_0^1  dt  \frac{(1-t)^{N-1}}{(N-1)!}
(*\exp A_n t - *\exp At) = 0.
\l{k24}
\ee
One has
$$
*\exp A_nt - *\exp At = \int_0^t d\tau *\exp A(t-\tau) * (A_n-A) *\exp
A_n\tau.
$$
Making use of lemma A.23, we obtain then estimation \r{k24}.

\subsection{Estimations for the commutator}

Let $\hat{A}=f(\hat{x})$,  $\hat{B} = g(\hat{k})$.  To investigate the
properties of the commutator $\hat{K} =  [\hat{A};  \hat{B}]$,  it  is
convenient to  introduce  the notion of $\hat{x}\hat{k}$-symbol of the
operator instead       of       Weyl       symbol.       For       the
$\hat{x}\hat{k}$-quantization, the     operator
$e^{i\beta\hat{x}}e^{i\alpha\hat{k}}$
corresponds to the function $e^{i\beta {x}}e^{i\alpha {k}}$.
Therefore, the function
$$
A(x,k) =  \int  d\alpha  d\beta \tilde{A}(\alpha,\beta) e^{i\alpha k +
i\beta x}
$$
corresponds to the operator
$$
\hat{A} =  \int  d\alpha  d\beta  \tilde{A}(\alpha,\beta)
e^{i\beta \hat{x}} e^{i\alpha \hat{k}}
$$
For $\hat{x}  \hat{k}$-quantization,  the  *-product  defined from the
relations $\hat{C} = \hat{A} \hat{B}$, $C=A*B$ has the form \c{M1,KM}
$$
(A*B)(x,k) = A(x, k - i \frac{\partial}{\partial y}) B(y,k) |_{y=x}.
$$

{\bf Lemma A.27.} {\it
1. Let $A(x,k) = \varphi_1(x)  \varphi_2(k)$  with  bounded  functions
$\varphi_1$, $\varphi_2$. Then $||\hat{A}|| < \infty$. \\
2. Let $A \in L^2({\bf R}^{2d})$.  Then $\hat{A}$ is a Hilbert-Schmidt
operator.
}

{\bf Proof.}    1.    One     has     $\hat{A}=     \varphi_1(\hat{x})
\varphi_2(\hat{k})$, $||\hat{A}||      \le      ||\varphi_1(\hat{x})||
||\varphi_2(\hat{k})|| = \max |\varphi_1| \max |\varphi_2| < \infty$.\\
2. One has
$$
Tr A^+A = \frac{1}{(2\pi)^{2d}} \int dx dk |A(x,k)|^2 < \infty.
$$
The commutator $\hat{K} = [f(\hat{x}),  g(\hat{k})]$ has the following
$\hat{x}\hat{k}$-symbol:
\beb
K(x,k) = [g(k) - g(k - i \frac{\partial}{\partial x}) ] f(x) =
\sum_{n=0}^L \frac{\partial^n   g}{\partial   k^{i_1}   ...   \partial
k^{i_n}} (-i)^n  \frac{\partial^n  f}{\partial  x^{i_1}  ...  \partial
x^{i_n}}  \crr -   \int_0^1   d\alpha   \frac{(1-\alpha)^L}{L!}   (-i)^{L+1}
\frac{\partial^{L+1} g(k-i \alpha \frac{\partial}{\partial x})}
{\partial k^{i_1}  ...  \partial  k^{i_{L+1}}  }  \frac{\partial^{L+1}
f}{\partial x^{i_1} ... \partial x^{i_{L+1}} }.
\eeb

{\bf Lemma A.28.} {\it Let
$C(x,k) = A(k-i\alpha \partial/\partial x) B(x) $. Then $||C||_{L^2} =
||A||_{L^2} ||B||_{L^2}$.
}

{\bf Proof.} Consider the Fourier transformation of the function $A$:
$$
A(k) = \int d\gamma \tilde{A}(\gamma) e^{i\gamma k}.
$$
One has $||A||_{L^2} = (2\pi)^{d/2} ||\tilde{A}||_{L^2}$ and
$$
C(x,k) = \int d\gamma \tilde{A}(\gamma) e^{i\gamma k}
e^{\gamma \alpha
\frac{\partial}{\partial x}} B(x).
$$
Since $e^{\gamma \alpha
\frac{\partial}{\partial x}} B(x) = B(x+\gamma\alpha)$, one has
\beb
||C||_{L^2}^2 = \int dk dx d\gamma_1 d\gamma_2
\tilde{A}^*(\gamma_1) e^{-i\gamma_1k} B^*(x+\gamma_1\alpha)
\tilde{A}(\gamma_2) e^{i\gamma_2k} B(x+\gamma_2\alpha) = \crr
(2\pi)^d \int  d\gamma  |\tilde{A}(\gamma)|^2  \int   dx   |B(x+\gamma
\alpha)|^2 = ||A||_{L^2}^2 ||B||_{L_2}^2.
\eeb
Lemma A.28 is proved.

We have obtained the following important statement.

{\bf Lemma A.29.} {\it Let
$\frac{\partial^n f}{\partial x^{i_1} ... \partial x^{i_n}}$,
$\frac{\partial^n g}{\partial   k^{i_1}   ...  \partial  k^{i_n}}$  be
bounded functions, $m,n=\overline{1,L}$, while
$$
\frac{\partial^{L+1} f}{\partial x^{i_1} ...  \partial  x^{i_{L+1}  }}
\in L^2, \qquad
\frac{\partial^{L+1} g}{\partial k^{i_1} ...  \partial  k^{i_{L+1}  }}
\in L^2.
$$
Then $[f(\hat{x}), g(\hat{k})]$ is a bounded operator.
}

\subsection{Asymptotic expansions of Weyl symbol }

To check the property of  Poincare  invariance,  it  is  important  to
investigate the large-$k$ expansion of the Weyl symbols. Introduce the
correponding definitions.

{\bf Definition A.4.} {\it 1.  We say that a smooth function $A(x,n)$,
$x,n \in  {\bf  R}^d$,  $|n|<1$,  is  of  the  calss  $\cal  L$ if the
functions
\be
\frac{\partial^I}{\partial n_{i_1} ...  \partial n_{i_I}} x_{j_1}  ...
x_{j_J} \frac{\partial^M}{\partial x_{m_1} ... \partial x_{m_M} }A
\l{k26}
\ee
are bounded.\\
2. Let $A_s \in {\cal L}$, $s=\overline{1,\infty}$. We say that ${\cal
L}-\lim_{s\to\infty} A_s = 0$ if
$$
\sup_{|n|\le 1}
\lim_{s\to\infty}
\left| \frac{\partial^I}{\partial n_{i_1} ...  \partial n_{i_I}}
x_{j_1}  ... x_{j_J} \frac{\partial^M}{\partial x_{m_1}  ...
\partial  x_{m_M}  }A
\right| = 0.
$$
}

Definitions A.2 and A.4 imply the following statement.

{\bf Lemma  A.30.  }  {\it 1.  Let $A\in {\cal L}$.  Then the function
$B(x,k) = A(x,k/\omega_k)$ is of the class ${\cal B}_0$.\\
2. Let ${\cal L}-\lim_{s\to\infty} A_s = 0$. Then
${\cal B}_0-\lim_{s\to\infty} A_s(x,k/omega_k) = 0$.
}

Making use  of definition A.2 and lemma A.25,  we obtain the following
corollary.

{\bf Corollary.} {\it 1.  Let  $A\in  {\cal  L}$.  Then  the  function
$\omega_k^{-\alpha} A(x,k/\omega_k)$    is   of   the   class   ${\cal
A}_{\alpha}$.\\
2. Let ${\cal L}-\lim_{s\to\infty} A_s = 0$. Then
${\cal A}_{\alpha}-\lim_{s\to\infty} \omega_k^{-\alpha}
A_s(x,k/omega_k) = 0$.
}

{\bf Definition A.5.} {\it 1.  A formal asymptotic expansion is a  set
$\check{A}$ of  $\alpha  \in  {\bf R}$ and functions $A_0,A_1,...  \in
{\cal L}$.
We say that the formal asymptotic expansions
$\check{A} = (\alpha, A_0,A_1,..)$ and
$\check{B} = (\beta, B_0,B_1,..)$ are equivalent if
$\alpha-\beta$ is an integer number and $A_{l-\alpha+\beta} = B_l$ for
all $l=\overline{-\infty,+\infty}$  (we assume $A_l = 0$ and $B_l = 0$
for $l<0$. We denote formal asymptotic expansions of Weyl symbols as
$$
\check{A} \equiv   \sum_{n=0}^{\infty}   \omega_k^{-n-\alpha}   A_n(x,
k/\omega_k).
$$
If $A_0=0$,  ..., $A_{l-1}=0$, $A_l\ne 0$, the quantity $deg \check{A}
\equiv \alpha + n$ is called  as  a  degree  of  a  formal  asymptotic
expansion $\check{A}$.
\\
2. Let  $\check{A}_s$,  $s=\overline{1,\infty}$  and  $\check{A}$   be
formal asymptotic expansions of Weyl symbols. We say that
$F.E-\lim_{s\to\infty} \check{A}_s   =  A$  if  $\alpha_s=\alpha$  and
${\cal L}-\lim_{s\to\infty} (A_{s,n} - A_s) = 0$.
}

The summation and multiplication by numbers are obviously defined:
$$
\check{A}+ \lambda        \check{B}        =       \sum_{n=0}^{\infty}
\omega_k^{-n-\alpha} (A_n(x, k\omega_k) + \lambda B_n(x,k/\omega_k)).
$$
The product of formal asymptotic expansions of Weyl symbols
$$
\check{A} \equiv   \sum_{n=0}^{\infty}   \omega_k^{-n-\alpha}   A_n(x,
k/\omega_k),
\qquad
\check{B} \equiv   \sum_{n=0}^{\infty}   \omega_k^{-n-\beta}
B_n(x, k/\omega_k)
$$
is defined as
$$
\check{A}\check{B} \equiv                          \sum_{n=0}^{\infty}
\omega_k^{-n-\alpha-\beta} \sum_{s,l \ge 0; s+l = n}
A_s(x, k/\omega_k) B_l(x, k/\omega_k).
$$
Let $f=f(x)$, $f \in {\cal C}$. Then
$$
f(x) \check{A} \equiv
\sum_{n=0}^{\infty}   \omega_k^{-n-\alpha}
f(x) A_n(x,k/\omega_k).
$$
One also defines
$$
\omega_k^{-\beta} \check{A} \equiv
\sum_{n=0}^{\infty}   \omega_k^{-n-\alpha-\beta}
A_n(x,k/\omega_k)
$$
and
$$
\frac{\partial \check{A}}{\partial k_s} =
\sum_{l=0}^{\infty}   \omega_k^{-l-\alpha-1}
\left[
- (l+\alpha)   A_l(x,n)  +  \frac{\partial  A_l}{\partial  n_p}  (x,n)
(\delta_{ps}- n_pn_s) \right]|_{n=k/\omega_k}
$$
The *-product of formal asymptotic expansions is introduced as
\beb
\check{A} * \check{B} \equiv \sum_{K=0}^{\infty} \sum_{n_1n_2  \ge  0,
n_1 + n_2 = K} \frac{i^{n_1-n_2} }{n_1! n_2! 2^{n_1+n_2}}
\frac{\partial^{n_1+n_2}}{\partial x^{i_1} ... \partial x^{i_{n_2}}
\partial k^{j_1} ... \partial k^{j_{n_1}} }
\sum_{l_1=0}^{\infty} \omega_k^{-l_1-\alpha_1} A_{l_1} (x, k/\omega_k)
\crr \times
\frac{\partial^{n_1+n_2}}{\partial x^{j_1} ... \partial x^{j_{n_1}}
\partial k^{i_1} ... \partial k^{i_{n_2}} }
\sum_{l_2=0}^{\infty} \omega_k^{-l_2-\alpha_2} A_{l_2} (x, k/\omega_k)
\eeb
The formal  asymptotic  expansions  $\check{A}  *  \omega_k^{\alpha}$,
$\check{A} * f(x)$ are defined analogously. The *-exponent of a formal
asymptotic expansion $\check{A}$ is defined as
$$
*\exp \check{A} - 1 = \sum_{n=1}^{\infty} \frac{\check{A}^{*n}}{n!}
$$
provided that $deg A$ is a positive integer number.

{\bf Definition A.6.} {\it
1. An asymptotic expansion of the Weyl symbol is a set  $\underline{A}
\equiv (A,\check{A})$  of  the Weyl symbol $A$ and a formal asymptotic
expansion $\check{A}$ such that
$$
A(x,k) - \sum_{l=0}^{n-1}
\frac{A_l(x,k/\omega_k)}{\omega_k^{l+\alpha}} \in {\cal A}_{n+\alpha}
$$
for all $n=\overline{0,\infty}$.\\
2. We say that $E-\lim_{s\to\infty} \underline{A_s} = \underline{A}$ if
$F.E-\lim_{s\to\infty} \check{A}_s = \check{A}$ and
$$
{\cal A}_{n+\alpha} -\lim_{s\to\infty}
( A_s(x,k) - \sum_{l=0}^{n-1}
\frac{A_{s,l}(x,k/\omega_k)}{\omega_k^{l+\alpha}})
=
A(x,k) - \sum_{l=0}^{n-1}
\frac{A_l(x,k/\omega_k)}{\omega_k^{l+\alpha}}
$$
for all $n=\overline{0,\infty}$.
}

{\bf Remark.} For given Weyl symbol $A$,  the asymptotic expansion  is
not unique. For example, let
$$
A(x,k) = m^2f(x)/\omega_k.
$$
One can choose $\alpha=2$, $A_0(x,n) = m^2f(x)$ anf find
$A(x,k) = \omega_k^{-2} A_0(x,k/\omega_k)$. On the other hand, one can
set $\alpha=0$,  $A_0(x,n)  =  f(x)  (1-n_in_i)$  and obtain $A(x,k) =
A_0(x,k/\omega_k)$ since $\omega_k^2- k_ik_i = m^2$.  We  see  that  a
degree is  a  characteristic  feature of an expansion rather than of a
symbol.

Let $\underline{A}=  (A,\check{A})$,  $\underline{B} = (B,\check{B})$.
Denote
$\underline{A}  *   \underline{B}   \equiv   (A*B,   \check{A}*
\check{B})$,\\
$\omega^{\alpha}_k * \underline{A} \equiv   (\omega_k^{\alpha} * A,
\omega_k^{\alpha} * \check{A})$,\\
$f(x)* \underline{A}   \equiv   (f(x) * A,  f(x) *   \check{A} )$,\\
$*\exp \underline{A} - 1 \equiv (*\exp A - 1, * \exp \check{A} - 1)$.

Theorems A.14, A.18, A.21 and lemmas A.25 and A.26 imply the following
statements.

{\bf Theorem A.31.} {\it
1. Let $\underline{A}$be an asymptotic expansion  of  a  Weyl  symbol.
Then $\omega^{\alpha}_k  *  \underline{A}$  and $f(x) * \underline{A}$
are asymptotic expansions of Weyl symbols under conditions of  theorem
A.14, while  $*\exp \underline{A} - 1$ is an asymptotic expansion of a
Weyl symbol,  provided that $deg  \check{A}$  is  a  positive  integer
number.\\
2. Let $\underline{A}$ and $\underline{B}$ be asymptotic expansions of
Weyl symbols.  Then  $\underline{A}  * \underline{B}$ is an asymptotic
expansion af a Weyl symbol.
}

{\bf Theorem A.32.} {\it
1. Let $E-\lim_{n\to\infty} \underline{A}_n = \underline{A}$. Then: \\
(a) $E-\lim_{n\to\infty} \omega_k^{\alpha} * \underline{A}_n
\omega_k^{\alpha} * \underline{A}$; \\
(b) $E-\lim_{n\to\infty} f(x) * \underline{A}_n
f(x) * \underline{A}$ under conditions of theorem A.14; \\
(c) $E-\lim{n\to\infty}   (*\exp   \underline{A_n}   -  1)  =  *  \exp
\underline{A} -1$ if $deg \check{A}_n$,  $deg \check{A}$ are  positive
integer numbers.\\
2. Let $E-\lim_{n\to\infty} \underline{A}_n = \underline{A}$ and
$E-\lim_{n\to\infty} \underline{B}_n = \underline{B}$. Then
$E-\lim_{n\to\infty} \underline{A}_n * \underline{B}_n =  \underline{A}
* \underline{B}$.
}

The time derivative of  teh  asymptotic  expansion  $\underline{A}(t)$
with respect to $t$ is defined in a standard way
$$
E-\lim_{\delta t \to 0 } \frac{\underline{A}(t+\delta t) -
\underline{A}(t)}{\delta t} = \frac{d\underline{A}(t)}{dt}.
$$
The integral $\int_{t_1}^{t_2} \underline{A}(t) dt$ is also defined in
a standard way.

Theorem A.32 imply the following statement.

{\bf Theorem  A.33.}  {\it 1.  Let $\underline{A}(t)$ be a continously
differentiable asymptotic expansion of a Weyl symbol. Then \\
(a)
$\frac{d}{dt} (\omega_k^{\alpha} * \underline{A}) = \omega_k^{\alpha} *
\frac{d\underline{A}}{dt}$; \\
(b)
$\frac{d}{dt} (f(x) * \underline{A}) = f(x) *
\frac{d\underline{A}}{dt}$ under conditions of theorem A.14. \\
(c)
$\frac{d}{dt} (*\exp   \underline{A}   -   1)   =    \int_0^1    d\tau
e^{\underline{A} (t-\tau)}      *      \frac{d\underline{A}}{dt}     *
e^{\underline{A} \tau}$;\\
(d) $\frac{d}{dt}      (\underline{A}      *      \underline{B})     =
\frac{d}{dt}\underline{A} *   \underline{B}    +    \underline{A}    *
\frac{d}{dt}\underline{B}$.
}

The only nontrivial statement is  (c).  It  is  proved  by  using  the
identity \c{KM}
$$
*\exp \underline{A}_1  -  *\exp \underline{A}_2 = \int_0^1 d\tau *\exp
(\underline{A}_1(1-\tau)) *  (\underline{A}_1  -  \underline{A}_2)   *
\exp(\underline{A}_2\tau).
$$

\newpage
\pagestyle{empty}

\end{document}